\documentstyle[psfig,mncite,subfigure]{mn}

\newcommand{\rmsub}[2]{#1_{\rm #2}} 
\newcommand{\bvec}[1]{\mbox{\boldmath ${#1}$}}

\begin{document}
 

\title[A search for starlight reflected from $\upsilon$~And's innermost 
planet]
{A search for starlight reflected from $\upsilon$~And's innermost 
planet}

\author
[A.Collier~Cameron, K. Horne, A. Penny and C. Leigh]
{Andrew Collier Cameron$^1$
\thanks{E-mail: andrew.cameron@st-and.ac.uk},
Keith Horne $^1$, 
Alan Penny$^2$ and 
Christopher Leigh$^1$
\\
$^1$School of Physics and Astronomy, Univ.\ of St~Andrews, 
St~Andrews, 
Scotland KY16 9SS \\
$^2$Rutherford Appleton Laboratory, Chilton, Didcot OX11 0QX, United 
Kingdom
\\
} 

\date{Received; accepted 2001}

\maketitle

\begin{abstract} 
In data from three clear nights of a WHT/UES run in 2000 Oct/Nov, and
using improved Doppler tomographic signal-analysis techniques, we have
carried out a deep search for starlight reflected from the innermost
of $\upsilon$ (upsilon) And's three planets.  We place upper limits on
the planet's radius $R_p$ as functions of its projected orbital
velocity $K_p \approx 139 \sin{i}$ km~s$^{-1}$ for various assumptions
about the wavelength-dependent geometric albedo spectrum $p(\lambda)$
of its atmosphere.  For a grey albedo $p$ we find $R_p \sqrt{p} <
0.98~R_J$ with 0.1\% false-alarm probability (4-$\sigma$).  For a
\scite{sudarsky2000albedo} Class~V model atmosphere, the mean albedo
in our 380-676~nm bandpass is $\left<p\right> \sim 0.42$, requiring
$R_p < 1.51~R_J$, while an (isolated) Class~IV model with
$\left<p\right> \sim 0.19$ requires $R_p < 2.23~R_J$.  The star's
$v_{rot}\sin{i}\sim 10$~km~s$^{-1}$ and estimated rotation period
$P_{rot} \sim 10$d suggest a high orbital inclination $i \sim
70-80^\circ$. We also develop methods for assessing the false-alarm 
probabilities of faint candidate detections, and for extracting 
information about the albedo spectrum and other planetary parameters
from faint reflected-light signals.
\end{abstract}


\section{Introduction}

The discovery of ``hot Jupiters'' -- giant exoplanets in 3 to 5-day
orbits about their parent stars \cite{mayor95_51peg,marcy97rev} -- has
overturned conventional theories of planetary-system formation. 
Theorists working to explain the solar system had posited that
Jupiters could form only outside 5~AU, where planetesimals retain ice
mantles.  Inward migration was considered possible, but deemed
unlikely since the planets would then be swallowed up by the parent
star.  Hot Jupiters have now revived the idea of inward migration, but
require mechanisms that halt some of the planets near 0.05~AU.

Theoretical studies of the structure and evolution of exoplanets are
yielding testable predictions.  Gas-giant models yield planet radii
$R_p \sim 1.0-1.7~R_{Jup}$ for $1 < M_p/M_{Jup} < 10$
\cite{saumon96theory,guillot96cegp,burrows97nongray}.  Hot Jupiters
have fully-convective interiors, so the core entropy fixes the radius
at a given mass \cite{burrows2000}.  Isolated Jupiters will therefore
cool and contract to about Jupiter's size in a few Gyr, but strongly
irradiated giant planets find it harder to cool.  Consequently the
mass-radius-age relation for hot Jupiters is determined largely by (1)
the age at which they arrived in their present close orbits, and (2)
their atmospheric albedos, which control the atmospheric
temperature-pressure gradient and hence the rate of cooling
\cite{burrows2000}.  

The discovery of transits of the planet orbiting
HD 209458 \cite{charbonneau2000transit} provided conclusive proof of
the existence and gas-giant nature of the ``hot Jupiters". 
Ultra-precise HST/STIS photometry of HD 209458 b's transit profile
yields a radius about 1.35 $R_{Jup}$ \cite{brown2001}, although the
planet has a mass only 0.63 $M_{Jup}$.  The next crucial step will be
to measure albedos directly.

\begin{figure}
\begin{tabular}{r}	
\psfig{figure=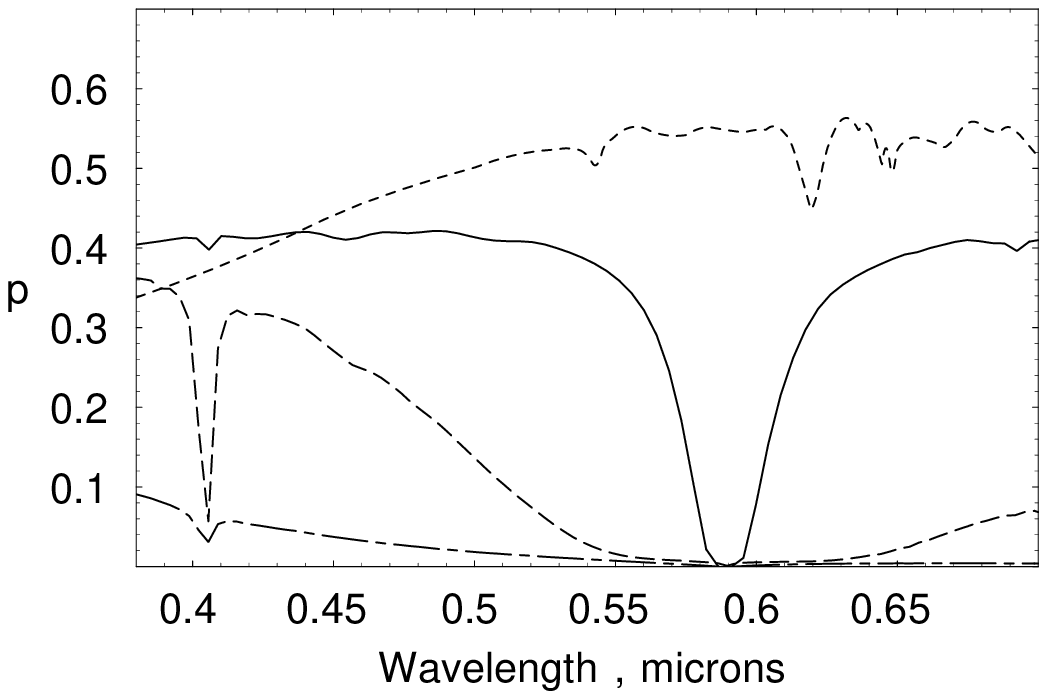,width=8.5cm} \\
\psfig{figure=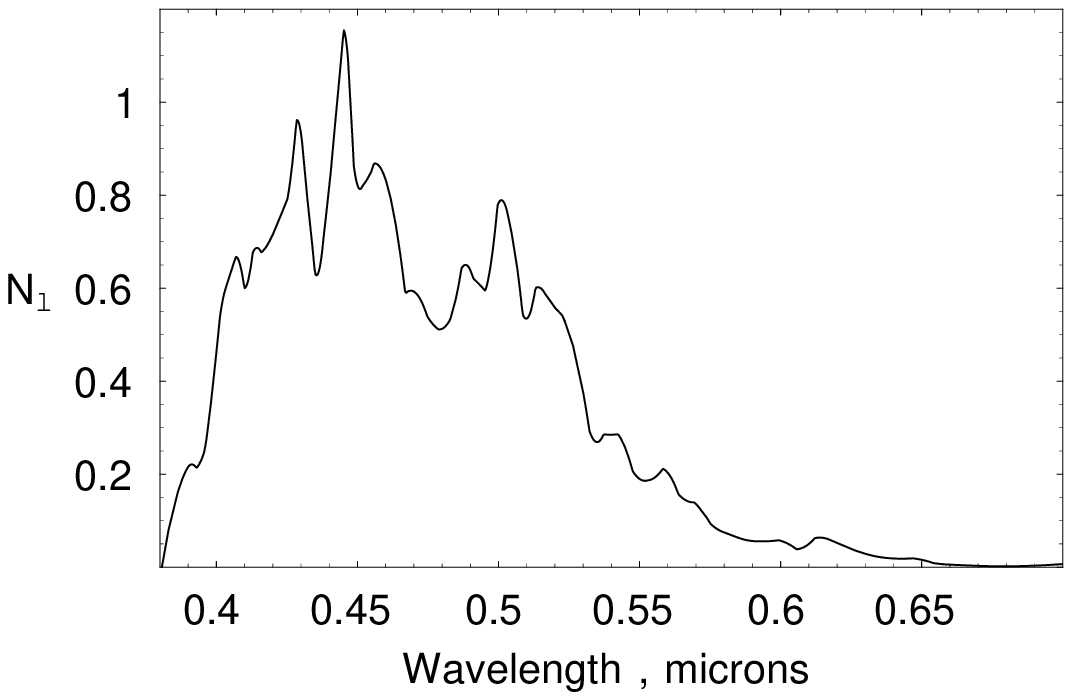,width=8.5cm}
\end{tabular}
\caption[]{Geometric albedo spectra for the Class V (solid line),
isolated (dashed) and irradiated (dot-dashed) Class IV models of
\scite{sudarsky2000albedo}, plotted over the wavelength range we
observed.  Together with a grey model having $p=0.42$, the Class V and
isolated Class IV models were used to probe the wavelength dependence
of candidate reflected-light signals.  The absorption features near
0.404 and 0.589 $\mu$m are due to potassium and sodium respectively. 
The geometric albedo spectrum of Jupiter (short dashes) is shown for
comparison, including methane absorption bands at 0.54 and 0.62
$\mu$m.  The lower panel shows the wavelength sensitivity of our
detection method, expressed as the deficit of detected photons
$N_{\lambda}$ per unit wavelength due to known absorption lines in the
stellar spectrum.  $N_{\lambda}$ is expressed in units of $10^{11}$
photons per $\mu$m, and gives the total photon deficit for a typical
group of 4 exposures.}
\label{fig:albsgeom}
\end{figure}

Spectral models\cite{seager98,marley99albedo,sudarsky2000albedo}
include thermal emission in the infrared and scattered starlight in
the optical.  The albedo is very sensitive to the depth of cloud formation. 
At the expected temperatures ($T_{\rm eff} \sim 1400$K) and gravities
($g\ge 36$ m s$^{-2}$) of massive planets such as $\tau$~Boo~b,
possible cloud condensates include upper cloud decks of MgSiO$_3$
(enstatite) and Al$_2$O$_3$ (corundum) at pressures $\sim 5$ to 10 bar,
and a deeper layer of Fe grains at pressures a factor 2 or so higher. 
If silicate cloud decks form this deep in the atmosphere or are
absent, much of the incident starlight is absorbed in
pressure-broadened absorption lines of neutral sodium and potassium,
leaving only a weak Rayleigh-scattered reflection signature at blue
wavelengths.  \scite{sudarsky2000albedo} predict this ``Class IV''
albedo spectrum for planets with high surface gravities.  At the lower
surface gravities $g\simeq 10$ m s$^{-2}$ expected of lower-mass
planets like $\upsilon$~And~b and HD 209458 b, ``Class V''
models predict silicate cloud decks forming higher in
the atmosphere at pressures around 0.3 bar, giving a geometric albedo
about 60\%\ that of Jupiter throughout most of the optical spectrum
(Fig.\ref{fig:albsgeom}).

Recent attempts to detect starlight reflected from $\tau$~Boo~b have
produced deep upper limits on the planet's albedo. 
\scite{charbonneau99tauboo} established an upper limit on the
fully-illuminated planet/star flux ratio $\epsilon_{0}\simeq 10^{-4}$. 
\scite{cameron99tauboo} claimed a detection with an average
$\epsilon_{0}=7.5\times10^{-5}$ from data secured in 1998 and 1999. 
The radius inferred for the planet was, however, implausibly large. 
Deeper observations made early in 2000 \cite{cameron2001tauboo} did
not confirm this candidate detection, but produced a more stringent
upper limit, $\epsilon_{0} < 3.5\times 10^{-5}$ between 387.4 and
586.3 nm.  The apparent tidal synchronisation of $\tau$ Boo with the
planet's orbit suggests a low inclination of order 40$^{\circ}$.  At
this inclination, the combined 1998-2000 WHT observations yielded a
99.9\% upper limit on the geometric albedo $p\le 0.22$, assuming a
planet radius $1.2 R_{Jup}$ as predicted by
\scite{burrows2000}.  This appears to rule out the presence of a high
cloud deck with the high reflectivity of the Class V models, which
would give $p\simeq 0.4$ over most of the visible spectrum.

Here we report results from a programme of observations designed to
detect the spectroscopic signature of starlight reflected from the
innermost of the three giant planets orbiting $\upsilon$~And
\cite{butler99upsand}.  In Section~\ref{sec:prior} we use the measured
system parameters to determine the {\em a priori} probabilities for
the planet's orbital velocity amplitude and the fraction of the star's
light that it intercepts.  In Section~\ref{sec:albedo} we discuss the
expected albedo and phase function that the planet might display.  In
Sections \ref{sec:obser} and \ref{sec:extraction} we describe the
acquisition and extraction of the echelle spectra.  The methods used
to remove the direct-starlight signature from the data, and to extract
the planetary signal by combining the profiles of the thousands of
photospheric lines recorded in each echellogram, are presented in
Section \ref{sec:aldecon}.  A matched-filter method for measuring the
strength of the reflected-light signal is described in
Section~\ref{sec:backpro}.  Finally in Section~\ref{sec:backpro} we
determine upper limits on the planet's radius for three different
models of the planet's albedo spectrum, and discuss the plausibility
of a candidate reflected-light signal that appears in the data at the
most probable velocity amplitude and signal strength.

\section{System parameters}
\label{sec:prior}

\begin{table*}
	\caption[]{Physical parameters for $\upsilon And$ and its innermost 
	planet.}
	\label{tab:params}
	\begin{tabular}{lcl}
		Parameter & Value (Uncertainty) & References \\
		           &        &  \\
		{\bf Star:}           &        &  \\
		Spectral type & F8V & \scite{gonzalez97,fuhrmann98,gonzalez98} \\
		$\rmsub{T}{eff}$ & 6100 (80) & \scite{gonzalez97,fuhrmann98,gonzalez98} \\
		$M_{\star}$ ($M_{\odot}$)  & 1.3 (0.1) & \scite{fuhrmann98,ford99} \\
		$R_{\star}$ ($R_{\odot}$) & 1.63 (0.06) & \scite{fuhrmann98,ford99} \\
		$\log g$ & 4.0 & \scite{gonzalez97,fuhrmann98,gonzalez98} \\
		$[$Fe/H$]$ & 0.09 to 0.17 & \scite{gonzalez97,fuhrmann98,gonzalez98} \\
		$\pi$ (mas) & 74.25 (0.70) & \scite{perryman97} \\
		$v\sin i$ (km~s$^{-1}$) & 9.5 (0.8) & \scite{fuhrmann98} \\
		$\log P_{rot}$ (days) & 1.08 (0.08) & \scite{henry2000,baliunas97,noyes84} \\
		Age (Gyr) & 2.6 to 4.8 & \scite{fuhrmann98,ford99} \\
		           &        &  \\
		{\bf Planet:} & & \\
		Orbital period $P_{orb}$ (days) & 4.61707 (0.00003) & Marcy, personal 
		communication \\
		Transit epoch $T_{0}$ (JD) & 2451821.457 (0.012) & Marcy, personal 
		communication \\
		$K_{\star}$ (m s$^{-1}$) & 74.5 (2.3) & \scite{butler99upsand} \\
		a (AU) & 0.059 (0.002) & \scite{butler99upsand} (revised value derived in this paper) \\
		$M_{p}\sin i$ ($M_{Jup}$) & 0.73 (0.04) & \scite{butler99upsand} (revised value derived in this paper) \\

	\end{tabular}
\end{table*}

$\upsilon$~And (HR 458, HD 9826) is a late-F main-sequence star with
parameters as listed in Table \ref{tab:params}.  
High-precision
radial-velocity studies have shown it to have a system of 3 planets
\cite{butler99upsand}.  For the purposes of this paper we are
concerned only with the innermost of these planets, whose properties
(as determined directly from radial-velocity studies or inferred using
the estimated stellar parameters) are also summarised in Table
\ref{tab:params}.  

Starlight reflected from the orbiting planet's atmosphere leaves a
detectable signature in observed spectra in the form of faint copies
of each of the stellar absorption lines, Doppler shifted by the
planet's orbital motion, and greatly diminished in brightness because
the planet intercepts only a fraction $(1/4)(R_p/a)^2 \sim 10^{-4}$ of
the starlight and only part of that is reflected back into space.

By detecting and characterizing the planetary reflected light signal,
we in effect observe the planet/star flux ratio $eps \equiv f_p/f_\star$
as a function of orbital phase $\phi$ and wavelength $\lambda$.
The information we aim to obtain (in order of increasing difficulty) 
comprises:
\begin{enumerate}
\item $K_p$, the planet's projected orbital velocity.
	From this we learn the orbital inclination $i$
	and hence the planet mass $M_p$,
	since $M_p\sin{i}$ is known from the star's Doppler wobble.
\item  $\epsilon_0$, the maximum strength of the reflected starlight.
	From this we constrain the planet's radius 
	since $\epsilon_0 = R_p/a \sqrt{p}$, 
	where $p$ is the geometric albedo.
\item $p(\lambda)$, the albedo spectrum, which depends on the
	composition and structure of the planetary atmosphere.
\item $g(\alpha,\lambda)$, the phase function describing the
	dependence of the amount of light reflected toward the observer
	on the star-planet-observer angle $\alpha$.
\item $\Delta$, the velocity width of lines in the reflected starlight.
	This depends mainly on the star's rotation in the frame of the orbiting
	planet, and to a lesser extent on the planet's rotation and surface winds.
\end{enumerate}
At this stage of the subject, our primary goal is to achieve
detections of the planetary reflected light signal, thereby reaching
step (2) in the above list -- measuring $K_p$ and $\epsilon_0$
and hence $R_p$ for an assumed albedo $p$.
With sufficient data we can measure $epsilon(\lambda)$
and hence $p(\lambda)$, but with present data the best we can do is
to adopt $p(\lambda)$ for various planetary atmospheric models
and for each one measure or place an upper limit on the planet
radius $R_p/a$.
To optimize the sensitivity for detection, we restrict observations
to gibbous phases of the planetary orbit, thus sacrificing
information about the phase function $g(\alpha,\lambda)$.
We elaborate these issues in more detail below.

\subsection{Radial velocity amplitude}

The radial velocity curve of the reflected light at orbital phase 
$\phi$ is
\begin{equation}
V_p(\phi)=K_{p}\sin 2\pi\phi,
\end{equation}
The orbital phase at time $t$ is given by $\phi=(t-T_{0})/P_{orb}$,
using the values of $P_{orb}$ and $T_{0}$ given in
Table~\ref{tab:params}.

The mass ratio $q=M_{p}/M_{\star}$ is determined from the observed
amplitude $K_{\star}$ of the star's reflex orbital velocity via:
\begin{eqnarray}
	\frac{q}{1+q}&=&\frac{K_{\star}}{\sin i}\frac{P_{orb}}{2\pi a}
	\nonumber\\
	&=&\frac{5.339\times 10^{-4}}{\sin i}
	 \left(\frac{K_{\star} }{ 74.5 \mbox{ m s}^{-1} } \right)
	 \left( \frac{ M_\star }{ 1.3 M_\odot } \right)^{-1/3}.
	\label{eq:qratio}
\end{eqnarray}

The apparent radial velocity amplitude $K_{p}$ of the
planet about the system's centre of mass is
\begin{equation}
	K_{p} = \frac{ 2 \pi a }{ P_{orb} } \frac{ \sin{i} }{ 1 + q }
       = 139 \frac{ \sin{i} }{ ( 1 + q )^{2/3} } 
       \left( \frac{ M_\star }{ 1.3 M_\odot } \right)^{1/3}
       {\rm km~s}^{-1}.
	\label{eq:kp}
\end{equation}
Kepler's third law gives
\begin{equation}
	a = 0.0592\left(\frac{M_{\star}}{1.3 M_{\odot}}\right)^{1/3}\mbox{ AU}.
	\label{eq:aorb}
\end{equation}

The orbital inclination $i$ is, according to the usual convention, the
angle between the orbital angular momentum vector and the line of
sight.  For all but the lowest inclinations, the planet's orbital
velocity amplitude is considerably greater than the typical widths of
the star's photospheric absorption lines.  Lines in the planet's
reflected-light spectrum should therefore be Doppler-shifted well
clear of their stellar counterparts, allowing clean spectral
separation for most of the orbit.

\subsection{Orbital inclination}

\begin{figure}
\psfig{figure=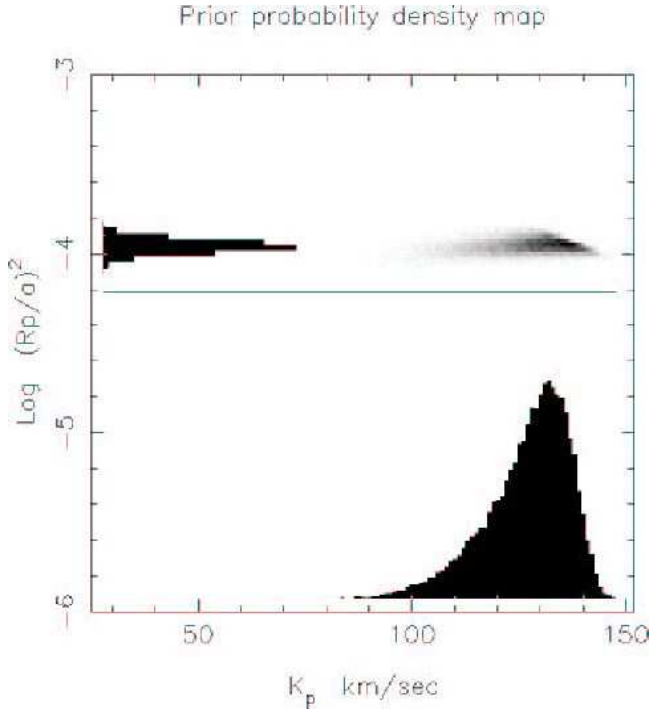,width=8.5cm}
\caption[]{The greyscale shows the prior joint probability density
function (PDF) for projected orbital velocity $K_{p}$ and the squared
ratio $(R_{p}/a)^{2}$ of the planet radius to the orbit radius, based
on the measured system parameters.  Darker shades in the greyscale
denote greater probabilities of the planet having the corresponding
combination of $(R_{p}/a)^{2}$ and $K_{p}$.  The one-dimensional
projections of the PDF on to the two axes are shown as histograms. 
They show that the planet is most likely to have $K_{p}\simeq 135$ km
s$^{-1}$ and $(R_{p}/a)^{2}\simeq 10^{-4}$. The horizontal line shows 
the value of $(R_{p}/a)^{2}$ for a 1 $R_{Jup}$ planet in the same orbit.}
\label{fig:prior}
\end{figure}

Orbital inclinations $i>80^{\circ}$ or so are ruled out by the absence
of transits in high-precision photometry \cite{henry2000}.  The
minimum inclination at which grazing transits occur is given by
\begin{equation}
	\cos i_{min}=\frac{R_{\star}+R_{p}}{a}.
	\label{eq:transit}
\end{equation}

Low inclinations are also ruled out by the star's chromospheric
activity levels.  The stellar radius and $v\sin i$ suggest an axial
rotation period of 9 days.  A $12\pm 2$ day period was estimated by
Baliunas et al.  (1997) based on four Ca~{\sc ii} H and K flux
measurements and the period-activity-colour relation of Noyes et al. 
(1984).  A more comprehensive study by \scite{henry2000}, based on an
additional 212 Ca~{\sc ii} H and K flux measurements in the 1996, 1997
and 1998 observing seasons shows the long-term average Ca~{\sc ii} H
and K flux level to be almost identical to the original estimate by
Baliunas et al.  The rms scatter in the 30-day means is 4.7\%\ of the
long-term mean flux level.  The 20\% uncertainty in the 11.7-day
period estimated from the long-term mean chromospheric flux is
therefore dominated by the intrinsic scatter in of about 0.08 dex in
the Noyes et al.  (1984) calibration.  Significantly faster axial
rotation is needed to match the observed $v\sin i$ at orbital
inclinations less than 60$^{\circ}$, but is hard to reconcile with the
observed chromospheric activity level.  Henry et al.  also point out
that frequency analyses of the short-term variability in the 1996 and
1998 seasons yielded possible low-amplitude rotational modulation
signals with periods of 11 and 19 days respectively, although both
were classified as ``weak''.

These rather loose constraints on $i$ require a Monte Carlo simulation
to express our knowledge of the system parameters.  For this we first
generate a distribution of $\sin i$ values
\begin{equation}
	\sin i = \frac{P_{CaII}.v\sin i}{2\pi R_{\star}}
\end{equation}
computed from randomly-chosen values of the stellar rotation period,
radius and $v\sin i$ and rejecting those combinations that yielded
$\sin i > 1$.  We assume gaussian distributions for the log of the
rotation period, the measured stellar mass and radius, and the stellar
reflex velocity $K_{\star}$.

High inclinations are then eliminated by the observed absence of
eclipses.  Low $i$ are then given a reduced probability consistent
with the measured $V_{rot} \sin{i} = 9.5$ km~s$^{-1}$ and estimate of
its rotation period $P_{rot} \sim 12\pm 2$d, which together favour a
relatively high $i$.  The lower histogram in Fig.~\ref{fig:prior}
shows the resulting probability distribution for $K_p$ (equivalent to
$i$).

\subsection{Rotational broadening}

\begin{figure}
\psfig{figure=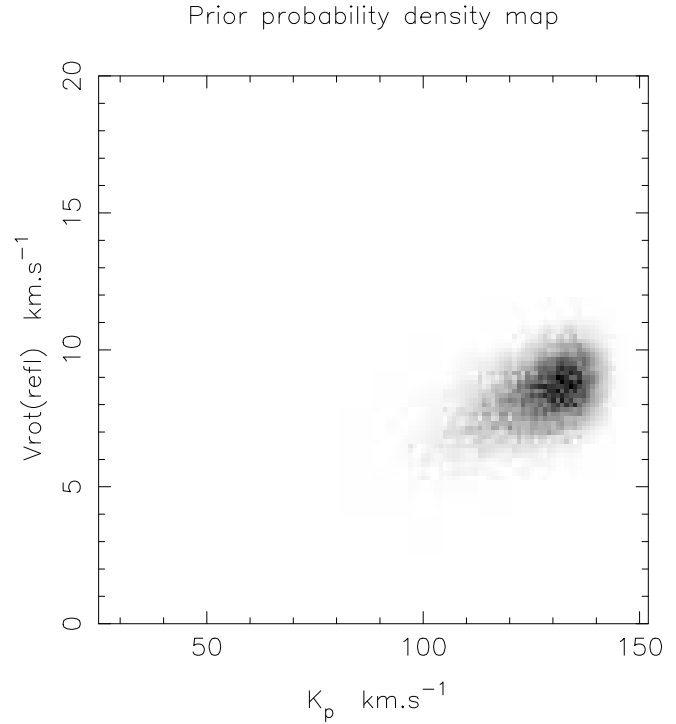,bbllx=106pt,bblly=73pt,bburx=481pt,bbury=404pt,angle=-90,width=8.5cm}
\caption[]{A Monte Carlo simulation shows how the apparent
stellar equatorial rotation speed $v_{e,refl}$ in the starlight
incident on the planet is expected to vary with the projected orbital
velocity amplitude $K_{p}$.  The system parameters are defined as in
Section~\ref{sec:prior}.  The median rotational broadening of the
reflected starlight is 8.4 km~s$^{-1}$, somewhat less than the $v\sin
i=9.5$ km~s$^{-1}$ of the direct starlight.}
\label{fig:verefl_kp}
\end{figure}

The matched-filter analysis used to search for the reflected-light
signal requires an {\em a priori} estimate of the rotational
broadening of the spectral-line profiles in starlight reflected from
the planet's atmosphere.

The Earth-bound observer sees stellar aborption lines that are
rotationally broadened by
\begin{equation}
	v_\star \sin{i} = \frac{ 2 \pi R_\star \sin{i} }{ P_{rot,*} }
		= 9.5\pm 0.8 {\rm km~s}^{-1}
\ .
\end{equation}
We assume that the planet orbits in the same direction as the stellar
rotation and that the orbital plane and stellar equator are
coincident.  The light received by the orbiting planet, and
subsequently reflected toward an observer in any direction, then has a
rotational broadening
\begin{equation}
	v_{refl} = 2 \pi R_\star 
		\left( \frac{1}{ P_{rot,*}} - \frac{1}{P_{orb}} \right)
\ .
\end{equation}
The Monte Carlo simulation gives the
resulting relationship between the broadening $v_{refl}$ of the
reflected starlight and $K_{p}$, as shown in Fig.~\ref{fig:verefl_kp}. 
The distribution of the predicted rotational broadening of the
reflected starlight has a mean 8.3 km~s$^{-1}$, $\sigma=1.1$ km
s$^{-1}$, and median 8.4 km~s$^{-1}$.  The planet line widths are
reduced at lower orbital inclinations, because the difference between
the orbital and stellar rotation frequencies is reduced.

\subsection{Albedo spectrum}
\label{sec:albedo}

The quantity that we observe is the ratio $f_{p}/f_{\star}$ of the
spectral-line strengths in the planet's light to those in the direct
stellar spectrum as a function of wavelength:
\begin{equation}
\epsilon(\alpha,\lambda)\equiv\frac{f_{p}(\alpha,\lambda)}{f_{\star}(\lambda)}
	=p(\lambda)g(\alpha,\lambda)\frac{R_{p}^{2}}{a^{2}}
    = \epsilon_{0}(\lambda)g(\alpha,\lambda).
\label{eq:fluxratio}
\end{equation}
This depends on the fraction $(R_{p}/2a)^{2}$ of the star's light
intercepted by the planet, and the fraction of this light that is
reflected towards the observer.  

The monochromatic stellar flux illuminating the planet, orbiting at
distance $a$, is
\begin{equation}
\rmsub{F}{incident}(\lambda)=\frac{L_{\star}(\lambda)}{4\pi a^{2}}.
\end{equation}

The geometric albedo $p(\lambda)$ is defined at $\alpha=0$
\begin{equation}
p(\lambda) = 
\frac{\rmsub{F}{reflected}(0,\lambda)}{\rmsub{F}{incident}(\lambda)}.
\end{equation}
Here
$\rmsub{F}{reflected}(0,\lambda)=\pi\rmsub{I}{reflected}(0,\lambda)$
is the disc-averaged flux of the starlight reflected from the planet
directly back toward the star, also measured at the planet's surface. 

As the planet orbits the star, the
observed flux varies with the star-planet-observer ``phase angle''
$\alpha$ as the product of the geometric albedo $p(\lambda)$ and a
``phase function'' $g(\alpha,\lambda)$ which is normalised to
$g(0,\lambda)=1$.

The flux received from the planet at Earth, is therefore
\begin{equation}
f_{p}(\alpha,\lambda) 
=p(\lambda)g(\alpha,\lambda)\rmsub{F}{incident}(\lambda)\frac{R_{p}^{2}}{D^{2}}.
\end{equation}

Since the stellar flux received at Earth, a distance $D$ away, is
\begin{equation}
f_{\star}(\lambda) = \frac{L_{\star}(\lambda)}{4\pi D^{2}},
\end{equation}
eq.~\ref{eq:fluxratio} follows.

\subsection{Phase function}

\begin{figure}
\psfig{figure=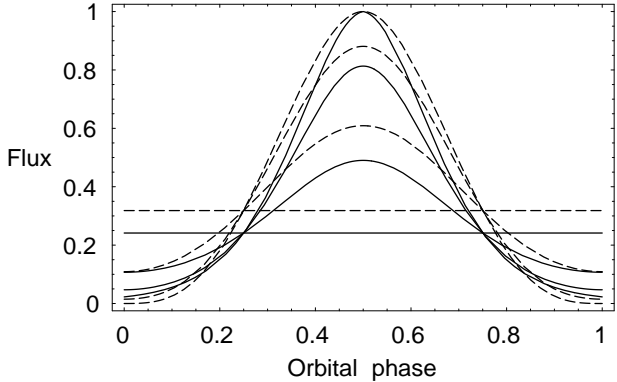,width=8.5cm} 
\caption[]{Brightness variation of model planet with orbital phase
assuming a Venus-like phase function (solid) or a Lambert sphere
(dashed) for orbit inclinations $i=0, 30, 60, 90^\circ$.  In both
cases, flux is measured relative to the value for illumination phase
angle $\alpha=0$.  The horizontal lines show the $i=0^\circ$ case,
while $i=90^\circ$ gives the greatest amplitude.  }
\label{fig:phasevar}
\end{figure}

The phase angle $\alpha$ depends on the orbital inclination $i$ and
the orbital phase $\phi$ (measured from transit or inferior
conjunction):
\begin{equation}
\cos\alpha = - \sin i .\cos 2\pi\phi.
\label{eq:alpha}
\end{equation}

The observations measure $\epsilon(\alpha,\lambda)$ over some range of
orbital phases $\phi$ and hence phase angles $\alpha$.  However, the
signal-to-noise ratio and orbital phase coverage of the observations
is not yet adequate to define the shape of the phase function. 
Accordingly, current practice is to adopt a specific phase function in
order to express the results in terms of the planet/star flux ratio
that would be seen at phase angle zero:
\begin{equation}
\epsilon_{0}(\lambda)=p(\lambda)\frac{R_{p}^{2}}{a^{2}}.
\label{eq:eps0}
\end{equation}

Since $a$ is tightly constrained through Kepler's third law, given the
period $P$ and the star mass $M_{\star}$, the measurements of
$\epsilon_{0}(\lambda)$ measure the product $R_{p}\sqrt{p(\lambda)}$.

In this paper we adopt an empirical phase function similar to that of
Venus and Jupiter.  The expressions given in
Appendix~\ref{sec:phasefuncs} (eqs.~\ref{eq:g_lambert} and and
Eq.~\ref{eq:g_venus} for the phase function of a Lambert sphere and
Venus respecively) yield the phase-dependent flux correction factors
plotted in Fig.~\ref{fig:phasevar}.

\subsection{Planet radius and surface gravity}

\begin{figure}
\psfig{figure=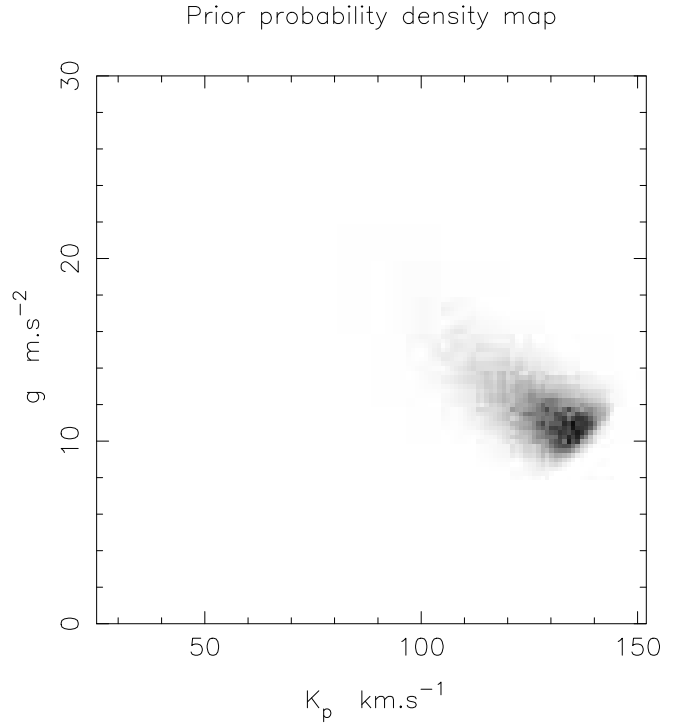,bbllx=106pt,bblly=73pt,bburx=481pt,bbury=404pt,angle=-90,width=8.5cm} 
\caption[]{A Monte Carlo simulation shows how the
gravitational acceleration $g$ at the planet's surface is expected to
vary with the projected orbital velocity amplitude $K_{p}$.  The
system parameters are defined as in Section~\ref{sec:prior}.  
For comparison, $g_{Jup}=26.5$ m~s$^{-2}$.}
\label{fig:grav_kp}
\end{figure}

Because the surface gravity appears to play an important role in
determining the height of the cloud deck and hence the albedo, we plot
the relationship between $g$ and $K_{p}$ derived from the Monte Carlo
simulations, in Fig.~\ref{fig:grav_kp}.  The planet's mass can be
determined once the orbital inclination is known.  Our a priori
knowledge of the radius, however, comes only from structural models
for close-orbiting giant planets.

In these models, the planet radius evolves with time and depends on
the planet mass.  For our purposes, theoretical mass-radius-age
relations define a range of plausible radii at each possible value of
the planet mass.  The range of possible planet radii was computed
specifically for $\upsilon$~And~b by \scite{guillot97abos}, allowing
for uncertainties in the orbital inclination and hence the planet's
mass.  Their radiative-convective gas giant models predict a radius
between 1.2 and 1.7 $\rmsub{R}{Jup}$ for $M_{p}=0.6 \rmsub{M}{Jup}$,
to between 1.15 and 1.3 $\rmsub{R}{Jup}$ for $M_{p}=1.15
\rmsub{M}{Jup}$.

When this mass-radius relation and its associated uncertainty are
incorporated in the Monte Carlo model, we find that as the inclination
decreases, the planet mass increases and the radius decreases.  The
lowest values of $g$ are therefore found when the orbit is, as is most
probable, nearly edge-on to the line of sight
(Fig.~\ref{fig:grav_kp}).  The median predicted value, $g = 11.6$
ms$^{-2}$, is close to the $\sim 10$ m s$^{-2}$ limit below which the
Class V albedo models of \scite{sudarsky2000albedo}) are applicable.

If Class V models apply, we can use them to estimate the planet's
brightness.  Viewed fully illuminated, the reflected light from a
planet with $(R_{p}/a)^{2}\simeq 10^{-4}$ and Class V albedo $p\simeq
0.42$ (cf.  the Class V models of \scite{sudarsky2000albedo}) is
expected to be 24000 times fainter than the direct stellar spectrum in
the visual band around 550 nm.  At a phase angle of 60$^{\circ}$,
however, the reflected spectrum will be a factor two fainter than this
(Fig.\ref{fig:phasevar}).

\section{Observations}
\label{sec:obser}

\begin{table*}
\caption{Journal of observations. The UTC mid-times and orbital 
phases are shown for the first and last groups of four spectra
secured on each night of observation. The number of groups is
given in the final column.}
\label{tab:journal}
\begin{tabular}{cccccc}
UTC start             & Phase   &  UTC End             & Phase  & 
Number\\
                      &         &                      &        & 
of groups\\
                      &         &                      &        &\\
2001 Oct 09  22:27:51 &	0.296	& 2001 Oct 10 06:20:26 & 0.366	&18\\
2001 Nov 06  19:43:04 & 0.335   & 2001 Nov 07 04:27:15 & 0.413  &16\\
2001 Nov 07  19:25:48 & 0.549   & 2001 Nov 08 04:29:30 & 0.630  &15\\
\end{tabular}
\end{table*}

The strong dependence of the flux ratio on phase angle indicates that
the best chance of a detection occurs shortly before and after
superior conjunction.  We observed $\upsilon$~And with the Utrecht
Echelle Spectrograph on the 4.2-m William Herschel Telescope at the
Roque de los Muchachos Observatory on La Palma, on the nights of 2000
Oct 10, Nov 6 and Nov 7.  These nights were chosen to cover orbital
phase ranges when the planet is on the far side of the star and
well-illuminated, and its spectral signature is Doppler-shifted well
clear of the wings of the stellar profile.  The third night was
partially affected by drifting cirrus cloud.  Two further nights were
also lost to cloud.

The detector was an array of two EEV CCDs, each with
$2048\times 4096$ 13.5-$\mu$m pixels.  The CCD was centred at 459.6 nm
in order 124 of the 31 g mm$^{-1}$ echelle grating, giving complete
wavelength coverage from 381.3 nm to 675.8 nm with minimal vignetting. 
The average pixel spacing was close to 1.6 km~s$^{-1}$, and the full
width at half maximum intensity of the thorium-argon arc calibration
spectra was 3.5 pixels, giving an effective resolving power $R=53000$.

Table \ref{tab:journal} lists the journal of observations for the
three nights that contributed to the analysis presented in this paper. 
The stellar spectra were exposed for between 300 and 500 seconds,
depending upon seeing, in order to expose the CCD to a peak count of
40000 ADU per pixel in the brightest parts of the image.  A 450-s
exposure yielded about $1.2\times 10^{6}$ electrons per pixel step in
wavelength in the brightest orders in typical (1 arcsec) seeing after
extraction.  We achieve this with the help of an autoguider procedure,
which improves efficiency in good seeing by trailing the stellar image
up and down the slit by $\pm 2$ arcsec during the exposure to
accumulate the maximum S:N per frame attainable without risk of
saturation.  Note that the 450-s exposure time compares favourably
with the 53-s readout time for the dual EEV CCD in terms of observing
efficiency -- the fraction of the time spent collecting photons is
above $90$ percent.  Following extraction, the S:N in the
continuum of the brightest orders is typically 1000 per pixel.

\section{Spectrum extraction}
\label{sec:extraction}

One-dimensional spectra were extracted from the CCD frames using an
automated pipeline reduction system built around the Starlink ECHOMOP
and FIGARO packages.  Nightly flat-field frames were summed from 50 to
100 frames taken at the start and end of each night, using an
algorithm that identified and rejected cosmic rays and other
non-repeatable defects by comparing successive frames.  The nightly
flat fields were then added to make a master flat-field for the entire
year's observations.

The initial tracing of the echelle orders on the CCD frames was
performed manually on the spectrum of $\upsilon$~And itself, using
exposures taken for this purpose without dithering the star up and
down the slit.  The automated extraction procedure then subtracted the
bias from each frame, cropped the frame, determined the form and
location of the stellar profile on each image relative to the trace,
subtracted a linear fit to the scattered-light background across the
spatial profile, and performed an optimal (profile and inverse
variance-weighted) extraction \cite{horne86extopt,marsh89optext} of
the orders across the full spatial extent of the object-plus-sky
region.  Flat-field balance factors were applied in the process.  In
all, 62 orders were extracted from each exposure.  The blue CCD
recorded orders 148 in the blue to 125 in the red, giving full
spectral coverage from 380.7 to 461.0 nm with considerable wavelength
overlap between adjacent orders.  Orders 122 to 85 were recorded on
the red CCD, covering the range 461.9 to 677.3 nm with good overlap. 
Orders 123 and 124 fell on the gap between the CCDs, but the loss in
wavelength coverage was minimal.

\section{Starlight-subtracted velocity-phase maps}
\label{sec:aldecon}

The maximum expected flux of starlight scattered from $\upsilon$~And
$b$ is, as we have seen, of order one part in 24000 of the flux
received directly from $\upsilon$~And itself.  In order to detect the
planet signal, we first subtract the direct stellar component from the
observed spectrum, leaving the planet signal embedded in the residual
noise pattern.  

To achieve this, we make a model of the direct starlight by aligning 
and summing all the spectra of the target from all nights of the run.
The contribution of the planet to this summed spectrum is blurred by 
the planet's orbital motion, so that to first order the planet's 
spectrum is eliminated from the composite ``template'' spectrum.
We then scale and distort this template spectrum to give the closest 
possible fit to each individual spectrum, using the spline-modulated 
Taylor-series method described in Appendix~\ref{sec:align}.

When the aligned template is subtracted from each spectrum, we find
that the residual spectrum contains a spatially fixed but temporally
varying pattern of ripples, superimposed on random noise.  The ripples
are attributable to a combination of time-dependent changes in the
detector's sensitivity, thermal flexure in the spectrograph causing
faint ghost reflections to shift slightly on the detector, and changes
in the strength and velocity of telluric-line absorption features in
some orders.  We map the form of the ripple pattern using
principal-component analysis, as described in
Appendix~\ref{sec:fpnoise}.  After subtracting this map, we are left
with a planet signal consisting of faint Doppler-shifted copies of
thousands of stellar absorption lines, deeply buried in noise.

The positions and identifications of most of these thousands of lines
are well-documented.  We use the Vienna atomic-line database (VALD;
\pcite{kupka99vald2}) to compile a list of the wavelengths and
relative central depths of the 3450 strongest lines falling in the
observed wavelength range, for a model atmosphere appropriate to the
spectral type and surface gravity of the star.  We then use
least-squares deconvolution (LSD; see Appendix~\ref{sec:lsd}) to
compute a composite profile which, when convolved with the line
pattern, yields an optimal fit to the residual noise spectrum.  This
procedure is similar to cross-correlation in terms of the gain in S:N
from the weighted summation of thousands of line profiles, but has the
additional advantage of eliminating sidelobes caused by crosstalk with
neighbouring lines.

\begin{figure}
\psfig{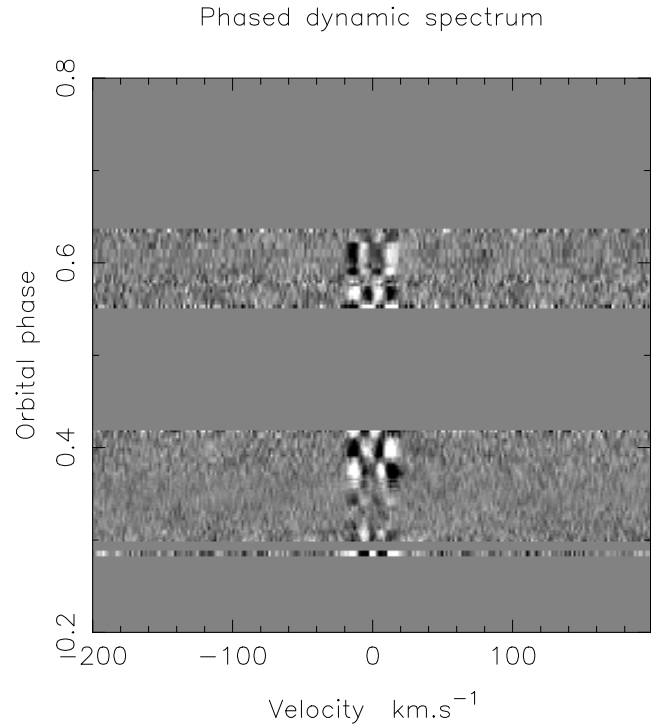} 
\caption[]{Velocity-phase map of deconvolved residual profiles derived
from WHT spectra secured on 2000 Oct 09, Nov 06 and Nov 07.  The line
weights in the deconvolution were defined assuming a grey albedo
spectrum.  The greyscale runs from black at $- 10^{-4}$ times the mean
stellar continuum level, to white at $+10^{-4}$.  The velocity scale
is in the reference frame of the star.  }
\label{fig:ph_obsgrey}
\end{figure}

The resulting devonvolved residual profiles are presented in greyscale
form as a two-dimensional ``velocity-phase'' map in
Fig.~\ref{fig:ph_obsgrey}.  This representation of the data shows a
strong pattern of distortions in the residual stellar profiles within
$\pm 20$ km s$^{-1}$ of the stellar velocity.  These undulations in
the deconvolved profiles appear to be caused by high-order mismatches
in the spectrum subtraction procedure.  Their amplitude is typically a
few parts in $10^{4}$ of the average continuum level.  They vary too
rapidly during the night to be attributable to, e.g. stellar surface
features causing time-dependent distortion of the stellar rotation
profile.  A more likely explanation is that the spatial profile
produced by the telescope dithering procedure was not exactly
repeatable from one frame to the next.  Given that the slit image is
slightly tilted with respect to the columns of the detector, this
could give rise to small changes in the detailed shape of the line
profile from one exposure to the next.  Fortunately these ripples only
affect a range of velocities at which the planet signature would in
any case be indistinguishable from that of the star.  We deliberately
avoided observing between phases 0.45 and 0.55 for this reason.

Outside the residual line profile, we would expect to see a planet
signature as a dark streak following a sinusoidal path that crosses
the profile from positive to negative velocity at phase 0.5.  No
obvious planet signature is, however, visible in
Fig.~\ref{fig:ph_obsgrey}.

\begin{figure}
\psfig{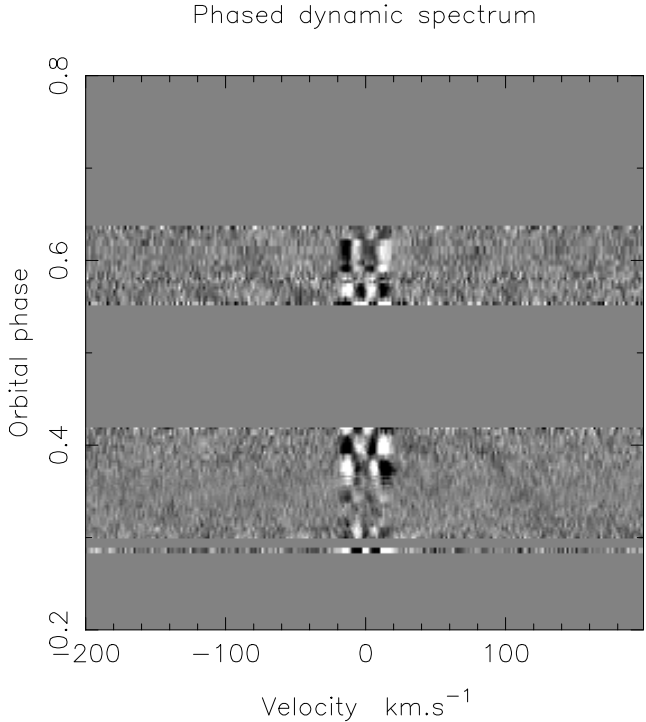}
\caption[]{As for Fig.~\ref{fig:ph_obsgrey}, with a simulated planet
signature added at an orbital inclination of 80$^\circ$.  The model
planet has a grey albedo spectrum with $p=0.42$ and a radius twice
that of Jupiter, and its signature appears as a dark streak crossing
from positive to negative velocity at phase 0.5.}
\label{fig:ph_sim80}
\end{figure}

We verified that a faint planetary signal is preserved through the
analysis in the presence of realistic noise levels, by adding a
simulated planetary signal to the observed spectra and performing the
same sequence of operations described above.  The simulation procedure
simply consisted of shifting and scaling the spectrum of
$\upsilon$~And according to the phase function, co-multiplying it by
an appropriate geometric albedo spectrum, and adding it to the
observed data.

To ensure a strong signal we assumed the planet to have a radius 2.0
times that of Jupiter, giving $(R_{p}/a)^{2}=2.57\times 10^{-4}$. 
Initially we used a wavelength-independent geometric albedo with
$p=0.42$, which was chosen to match the continuum albedo of a Class V
model unaffected by alkali-metal absorption.  When viewed at zero
phase angle, the planet-to-star flux ratio should thus be
$\epsilon_{0}=f_p/f_*=1.08\times 10^{-4}$.  We assumed an orbital
inclination of 80$^{\circ}$, giving an orbital velocity amplitude
$K_{p}=137$ km s$^{-1}$ and a rotational broadening of the reflected
starlight, $v_{e,refl}=8\pm 1$ km~s$^{-1}$.  We are therefore
justified in using the spectrum of $\upsilon$~And itself, without any
modification of the rotational broadening, to approximate the
reflected-light spectrum.

The injected planet signal is clearly visible as a dark streak in
Fig.~\ref{fig:ph_sim80}.  Given that no similar signal is easily seen
in Fig 6, we can conclude that the planet in $\upsilon$~And is fainter
than this one.

\section{Extracting the planet signature}
\label{sec:backpro}

The form of the expected planet signature in the velocity-phase
diagram can be represented accurately as a travelling Gaussian of
characteristic width $\Delta v_{p}$ whose velocity varies sinusoidally
around the orbit and whose strength is modulated according to the
phase function $g(\alpha,\lambda)$ (see
Appendix~\ref{sec:phasefuncs}).  The value of $\Delta v_{p}$ is
adjusted to match the expected rotational broadening of the reflected
starlight.  The amplitude $K_{p}$ of the velocity variation and the
form of the phase function both depend on the orbital inclination $i$.

\begin{figure}
\psfig{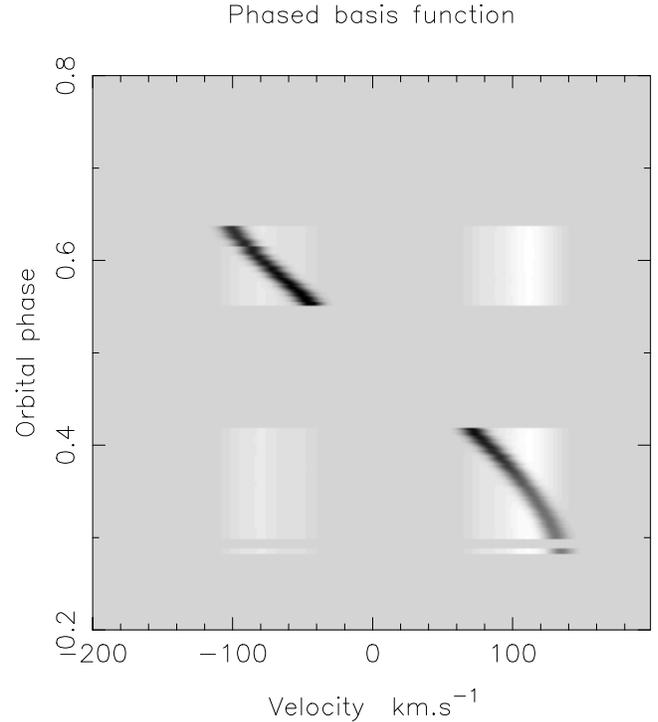} 
\caption[]{As for Fig.~\ref{fig:ph_obsgrey}, but showing the matched 
filter function used to measure $(R_{p}/a)^{2}$ at $K_{p}=137$ km~s$^{-1}$. 
The grey scale runs from -0.05 (black) to 0.01(white). Scaling this 
function to fit Fig.~\ref{fig:ph_obsgrey} (and others like it)
yields the strength of the planet signal for a given value of $K_{p}$.}
\label{fig:ph_basfunc}
\end{figure}

Fig.~\ref{fig:ph_basfunc} shows the form of this model signal at an
orbital inclination $i=80^{\circ}$.  The weakening of the simulated
planetary signature near quadrature (phases 0.25 and 0.75) is mostly
the effect of the phase function.  In some cases the signal may be
further attenuated near quadrature by the way in which the templates
are computed: since the planet signature is nearly stationary in this
part of the orbit, some of the signal will be removed along with the
stellar profile if many observations are made in this part of the
orbit.  The planet is detectable on a given night because its velocity
changes while those of the stellar lines do not.  Thus observations at
quadrature are less helpful.  In the present dataset, few observations
were made near quadrature so this problem does not arise.

The travelling gaussian in this image has nonetheless been modified to
give an even closer match to the observed planet signal, by
subtracting its own flux-weighted average.  This procedure mimics the
attenuation of the planet signal that occurs when the template
spectrum is subtracted from the individual spectra.  The main effect
of the template subtraction is to produce the bright vertical zones
seen centred at $v\simeq -70$ km s$^{-1}$ and $v\simeq +100$ km
s$^{-1}$ in Fig.~\ref{fig:ph_basfunc}.

\begin{figure}
\psfig{figure=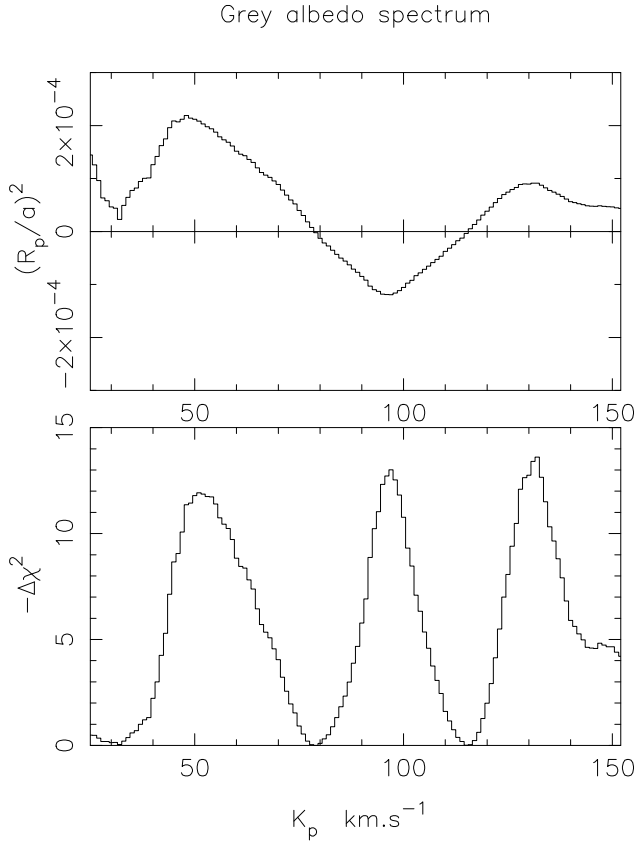,bbllx=7pt,bblly=0pt,bburx=350pt,bbury=453pt,width=8.5cm}
\caption{The upper panel shows the optimal scaling factor 
$(R_{p}/a)^{2}$ plotted against orbital 
velocity amplitude $K_{p}$, assuming a grey albedo spectrum with
$p=0.42$. The lower panel shows the associated
reduction in $\chi^{2}$ measured 
relative to the fit obtained in the absence of any planet signal, 
i.e. for $(R_{p}/a)^{2}=0$. 
Note that only positive values of $(R_{p}/a)^{2}$ are physically
plausible, so the middle peak in the $\Delta\chi^{2}$ plot is 
unambiguously a noise feature.
}
\label{fig:slice_grey}
\end{figure}

\begin{figure}
\psfig{figure=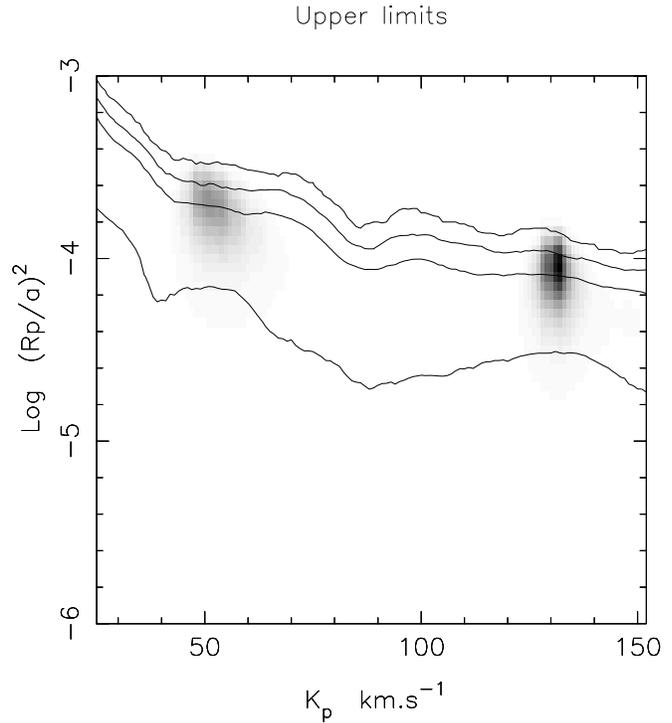,bbllx=106pt,bblly=73pt,bburx=481pt,bbury=404pt,angle=-90,width=8.5cm}
\caption{Relative probability map of model parameters $K_{p}$ and
$\log(\epsilon_{0}/p)=\log(R_{p}/a)^{2}$, derived from the WHT/UES
observations of $\upsilon$~And, assuming a grey albedo spectrum with
$p=0.42$.  The greyscale denotes the probability relative to the
best-fit model, increasing from 0 for white to 1 for black.  From
bottom to top, the contours show the 1, 2, 3 and 4$\sigma$ upper
limits limits on the signal strength derived from the bootstrap trials
The uppermost contour, for example, represents the value of
$\log(R_{p}/a)^{2}$ that was only exceeded in 3 of 3000 bootstrap
trials at each $K_{p}$.  It gives a robust empirical estimate of the
upper limit on the planet radius allowed by the data for the grey
albedo model with $p=0.42$ assumed here.  }
\label{fig:limits_obsgrey}
\end{figure}

\begin{figure}
\psfig{figure=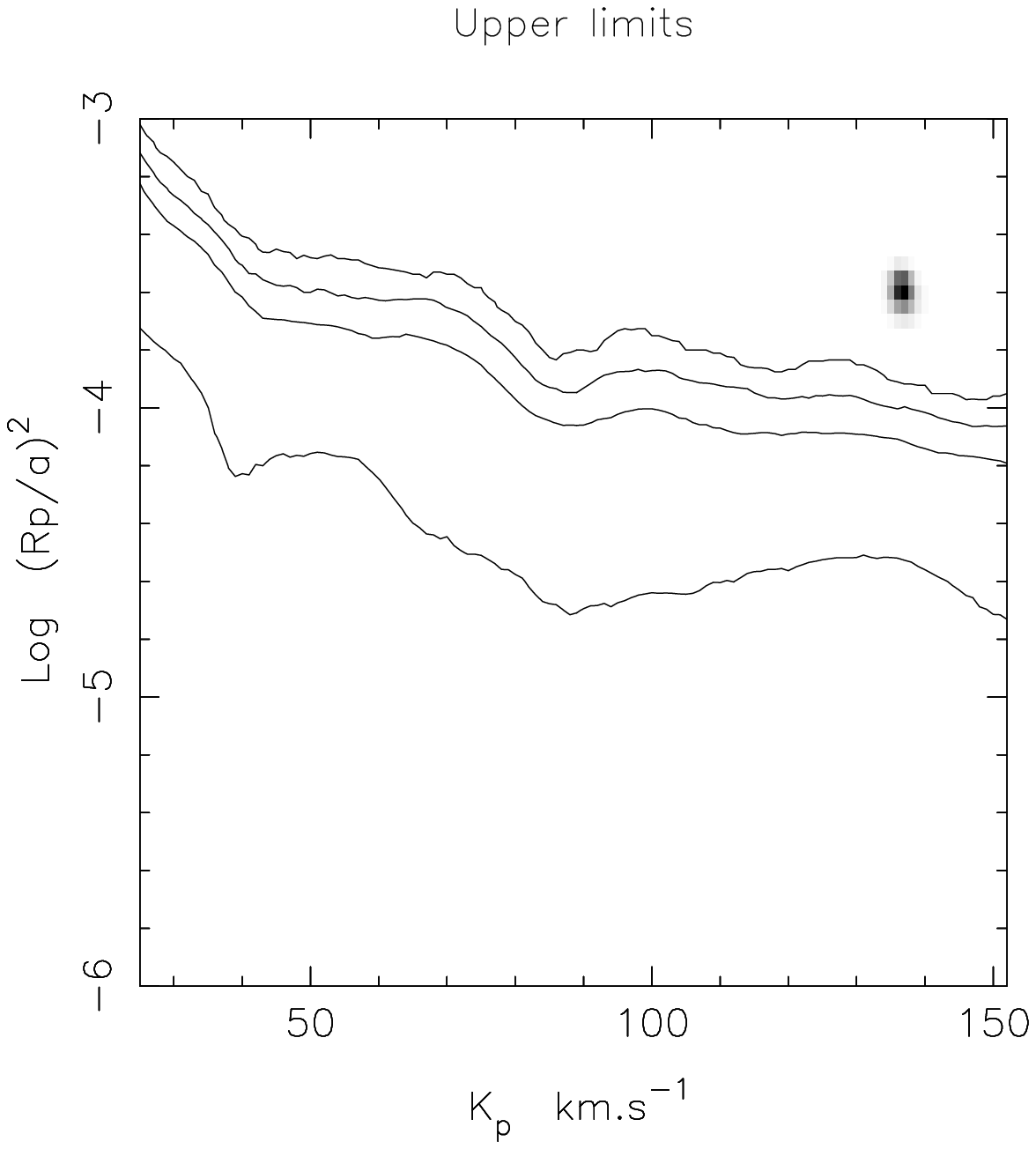,bbllx=106pt,bblly=73pt,bburx=481pt,bbury=404pt,angle=-90,width=8.5cm}
\caption{Relative probability map of model parameters $K_{p}$ and
$\log(\epsilon_{0}/p)=\log(R_{p}/a)^{2}$ for a simulated planet
signature with grey albedo $p=0.42$, $K_{p}=137$ km~s$^{-1}$, and
$R_{p}=2R_{Jup}$.  The greyscale and contours are defined as in
Fig.~\ref{fig:limits_obsgrey}.  The synthetic planet signature is
detected well above the 4$\sigma$ upper limit on the strength of noise
features in the absence of a planet signal.}
\label{fig:limits_difgrey}
\end{figure}

%
\section{Probability maps for $K_{p}$ and $R_{p}/a$}
\label{sec:results}

We use a sequence of such models for different values of $K_{p}$ as
matched filters to measure the strengths and velocity amplitudes of
possible faint planet signals of the expected form in the
velocity-phase maps.  At each of a sequence of trial values of $K_{p}$
we construct a model $H_{ij}(K_{p})$ like that shown in
Fig.~\ref{fig:ph_basfunc} and scale it to give an optimal fit to the
residual velocity-phase map $D_{ij}$ using the methods presented in
Appendix~\ref{sec:matchfilt}.  For an assumed albedo spectrum
$p(\lambda)$ and orbital velocity amplitude $K_{p}$, the fit of the
matched filter to the data measures $(R_{p}/a)^{2}$.

The badness-of-fit statistic $\chi^{2}$ gives a relative measure of
the probability that any signal detected could have arisen by chance,
and is quantified by
\begin{equation}
	\chi^{2}=\sum_{i,j}	
	\frac{(D_{ij}-(R_{p}/a)^{2}H_{ij}(K_{p}))^{2}}{\sigma^{2}_{ij}}.
\end{equation}
Here $\sigma^{2}_{ij}$ is the variance associated with $D_{ij}$, 
computed from the photon statistics of the original image and 
propagated through the deconvolution (see Appendix~\ref{sec:lsd}).

In Fig.~\ref{fig:slice_grey} we plot the optimum scale factor and the
associated improvement in $\chi^{2}$ measured relative to the
no-planet model.  For a noise-dominated signal, the estimated value of
$(R_{p}/a)^{2}$ will be negative about as often as it is positive, and
Fig.~\ref{fig:slice_grey} bears this out.  Three candidate peaks are seen
in the lower panel, of which only the first and the third correspond
to positive (and therefore physically plausible) reflected-light
fluxes.  The third (positive) peak gives the greatest improvement in
$\chi^{2}$ and is therefore the most probable, but only by a narrow
margin.

The relative probabilities of the fits to the data for different
values of the free parameters $R_{p}/a$ and $K_{p}$, given by 
\begin{equation}
	P(K_{p},(R_{p}/a)^{2})=\exp(-\chi^{2}/2),
\end{equation}
are shown in
Fig.~\ref{fig:limits_obsgrey}.  To make this figure, a planet signal
pattern as in Fig.~\ref{fig:ph_basfunc} was computed for many
different $K_{p}$, and each one was scaled by $(R_{p}/a)^{2}$ to fit
the residuals in Fig~\ref{fig:ph_obsgrey}.  The greyscale is defined
such that white represents the probability of a model fit where no
planet signal is present, while black represents the most probable
solution in the map.  The curves give 1, 2, 3 and 4$\sigma$ detection
thresholds derived from the bootstrap procedure discussed below.

To set an upper limit on the strength of the planet signal, or to
assess the likelihood that a candidate detection is spurious, we
need to compute the probability of obtaining such an improvement
in $\chi^{2}$ by chance alone.  In principle this could be done using
the $\chi^{2}$ distribution for 2 degrees of freedom.  In practice,
however, the distribution of pixel values in the deconvolved
difference profiles has extended non-gaussian tails that demand a
more cautious approach. 

Rather than relying solely on formal variances derived from photon
statistics, we use a ``bootstrap'' procedure to construct empirical
distributions for confidence testing, using the data themselves.  The
bootstrap procedure, detailed in Appendix E, interchanges at random
the order of the nights, and rearranges at random the order of spectra
in each night, thereby scrambling planet signals while retaining the
same pattern of systematic errors, phase sampling, and statistical
noise as in the actual data.

The 68.4\%, 95.4\%,99.0\% and 99.9\% percentage points of the
resulting bootstrap distribution of $(R_{p}/a)^{2}$ at each $K_{p}$
are shown as contours in Fig.~\ref{fig:limits_obsgrey} and
Fig.~\ref{fig:limits_difgrey}.  From bottom to top, the contours give
the 1, 2, 3, and 4$\sigma$ bootstrap upper limits on the strength of
the planet signal.  They represent the signal strengths at which
spurious detections occur with 32, 5, 1, and 0.1\% false alarm
probability respectively, for each fixed value of $K_p$.

\begin{table} 
	\caption[]{Upper limits on planet dimensions for various albedo
	models.  The upper limits are quoted for an assumed $K_{p}\simeq
	135$ km~s$^{-1}$, near the peak of the prior probability
	distribution for $K_{p}$.  Note that the results for the grey
	albedo model are quoted for unit geometric albedo.  To obtain the
	radii for grey models with other values of $p$, the radii in
	Column 4 should be divided by $\sqrt{p}$.}
	\label{tab:upperlim}
	\begin{tabular}{lccr} 
		Albedo & False-alarm  &  $(R_{p}/a)^{2}$ & $R_{p}/R_{Jup}$  \\
		model  & probability        &                  &  \\
		  	   & 			&				    &   \\
		Grey   &  0.1\% 	& 0.58E-04		& 0.98 \\
	($p=1.00$) &  1.0\% 	& 0.45E-04	    & 0.86 \\
		       &  4.6\% 	& 0.34E-04      & 0.75 \\
		  	   & 			&				    &  \\
	Class V    &  0.1\% 	& 1.42E-04		& 1.53 \\
	           &  1.0\% 	& 1.11E-04	    & 1.35 \\
		       &  4.6\% 	& 0.87E-04      & 1.19 \\
		  	   & 			&				    &  \\
	Class IV   &  0.1\% 	& 3.02E-04		& 2.23 \\
(Isolated)     &  1.0\% 	& 2.20E-04	    & 1.90 \\
		       &  4.6\% 	& 1.68E-04      & 1.66 \\
		  	   & 			&				    &  \\
	\end{tabular}
\end{table}


\subsection{Grey albedo model}

At the most probable values around $K_{p}\simeq 135$ km~s$^{-1}$, the
grey albedo model yields a 0.1\%\ bootstrap upper limit on the
planet/star flux ratio $\epsilon_{0}<5.84\times 10^{-5}$.  Adopting a
grey albedo model with unit geometric albedo, we find that the 0.1\%,
1.0\%\ and 4.6\%\ upper limits on $\epsilon_{0}$ at this velocity
correspond to upper limits on the planet radius as listed in
Table~\ref{tab:upperlim}. If the orbital inclination is lower, the
planet radius is less strongly constrained.

\begin{table*} 
	\caption[]{Orbital velocity amplitude, signal strength, planet
	radius, $\Delta\chi^{2}$ and false-alarm probabilities (FAP) for
	possible planet signals.  The FAP is computed for two different
	priors.  The first prior is uniform in $K_p$ from 44 to
	152~km~s$^{-1}$.  The second weights $K_p$ in proportion to the
	prior probability based on Ca{\sc ii} H \&\ K emission and $v\sin
	i$ constraints and the absence of eclipses.  In the latter case,
	the false-alarm probability for the $K_{p}=132$ km s$^{-1}$
	candidate is found to be between 9 and 10 percent.}
	\label{tab:fap}
	\begin{tabular}{lcccccc} 
		Albedo & $K_{p}$       &  $(R_{p}/a)^{2}$ &$R_{p}/R_{Jup}$&$\Delta\chi^{2}$ & FAP   & FAP \\
		model  & (km~s$^{-1}$) &  ($\times 10^{-4}$)	&     &  &(uniform weight) & ($K_{p}$ prior) \\
		  	   & 				&				    &     &                &       &       \\
		Grey   &  54 			&$0.87\pm 0.25$  	& $1.20\pm 0.17$ & 11.93          & 0.324 & 0.975 \\
	($p=1.00$) & 132 			&$0.39\pm 0.11$ 	& $0.80\pm 0.11$ & 13.61          &	0.259 & 0.094 \\
		  	   & 				&				    &     &                &       &       \\
	Class V    &  50 			&$2.08\pm 0.64$		& $1.85\pm 0.29$ & 10.39          & 0.423 & 0.983 \\
		       & 132 			&$1.09\pm 0.28$     & $1.34\pm 0.17$ & 14.61          & 0.240 & 0.087 \\
		  	   & 				&				    &      &                &       &       \\
    Class IV   &  49 			&$3.39\pm 1.31$	    & $2.36\pm 0.47$ &  6.66          & 0.534 & 0.995 \\
		       & 132 			&$1.73\pm 0.52$     & $1.68\pm 0.25$ & 10.78          & 0.285 & 0.099 \\
	\end{tabular}
\end{table*}

Two possible planet signals of comparable likelihood are seen.  Their
properties are listed in Table~\ref{tab:fap}.  The stronger, at
$K_{p}\simeq 132$ km~s$^{-1}$ ($i\simeq 80^{\circ}$), yields an
improvement $\Delta\chi^{2}=13.61$ over the model fit obtained
assuming no planet signal is present.  If this feature represents a
genuine planet signal, its velocity amplitude $K_{p}=132$ km~s$^{-1}$
implies a planet mass $M_{p}=0.74 M_{Jup}$ and (for $p=0.42$) a planet
radius $R_{p}=1.24\pm 0.17 R_{Jup}$.  The weaker candidate has
$K_{p}\simeq 54$ km~s$^{-1}$ ($i=22^{\circ}$) and
$\Delta\chi^{2}=11.86$, and gives a larger planet radius.

We used the bootstrap simulations to determine the probability that a
spurious detection with $\Delta\chi^{2}>13.61$ could be produced by a
chance alignment of noise features in the absence of a genuine planet
signal.  It is important to note that the location of the ``blob''
between the 2$\sigma$ and 3$\sigma$ bootstrap contours in
Fig.~\ref{fig:limits_obsgrey} does {\em not} imply a false-alarm
probability of $\sim 3$\%.  These contours give the false-alarm
probability only if the value of $K_{p}$ is known in advance, which is
not the case here.  The true false-alarm probability is greater, being
the frequency with which spurious peaks at {\em any} plausible value of
$K_{p}$ can exceed the $\Delta\chi^{2}$ of the candidate.  If we assume
that all values of $K_{p}$ are equally likely in the range 44
km~s$^{-1}<K_{p}<152$ km~s$^{-1}$ over which our phase coverage allows
signals to be detected, the false-alarm probability is found to be
26\%\ via the method described in Appendix~\ref{sec:fap}.

In practice, however, we are more likely to be fooled into believing
that a candidate detection near the peak of the {\em a priori}
probability distribution for $K_{p}$ is genuine, than would be the
case if the candidate appeared at a velocity that was physically
implausible given our existing knowledge of the system parameters.  We
can therefore refine the search range in $K_{p}$ using our knowledge
of the {\em a priori} probability distribution for $K_{p}$, via the
method presented in Appendix~\ref{sec:fap}.  The last two columns of
Table~\ref{tab:fap} show clearly that, while the unweighted false
alarm probabilities for the features near $K_{p}=132$ and $K_{p}=54$
km~s$^{-1}$ are comparable, the latter is almost certainly spurious. 
The false-alarm probability for the $K_{p}=132$ km~s$^{-1}$ feature
drops to 9.4\%\ when prior knowledge of $K_{p}$ is accounted for, while
that for the 54 km~s$^{-1}$ feature approaches unity.

\begin{figure}
\psfig{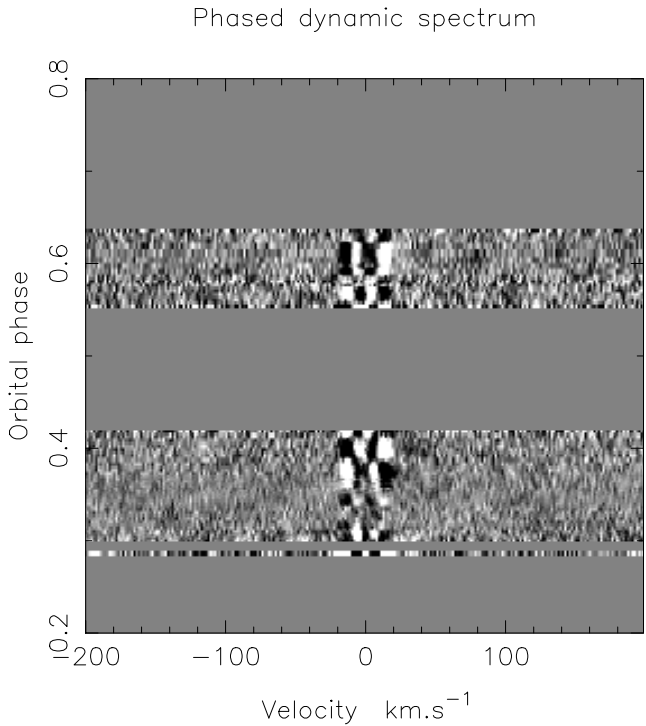} 
\caption[]{As for Fig.~\ref{fig:ph_obsgrey}, showing the time-series
of deconvolved profiles derived from the original observations 
assuming the albedo spectrum to be that of a Class V roaster.}
\label{fig:ph_obs_V}
\end{figure}

\begin{figure}
\psfig{figure=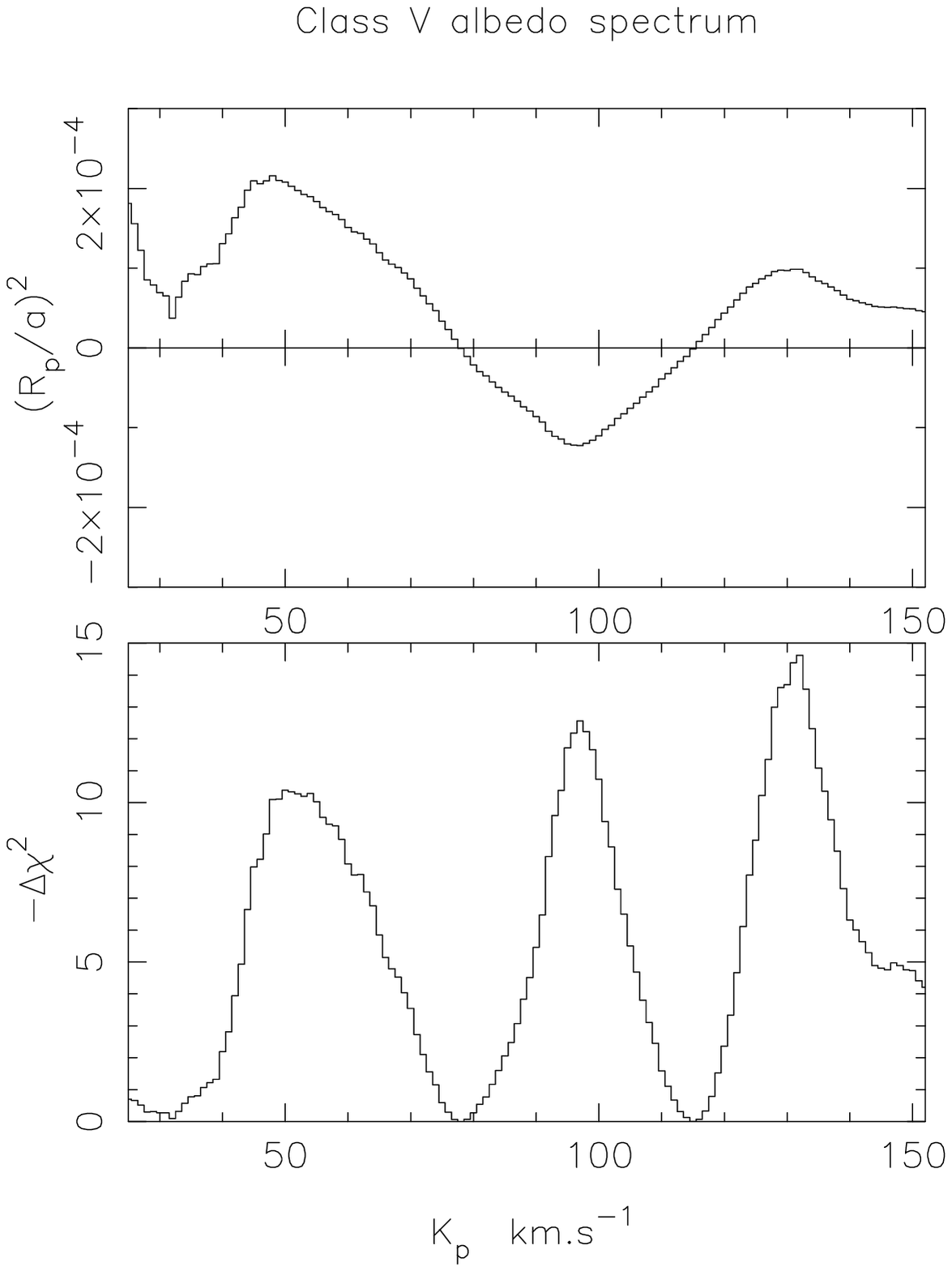,bbllx=7pt,bblly=0pt,bburx=350pt,bbury=453pt,width=8.5cm}
\caption{As for Fig.\ref{fig:slice_grey}, showing the 
optimal scaling factor 
$(R_{p}/a)^{2}$ and the associated
improvement in $\chi^{2}$ plotted against orbital 
velocity amplitude $K_{p}$, assuming a Class V albedo spectrum.
}
\label{fig:slice_V}
\end{figure}

\begin{figure}
\psfig{figure=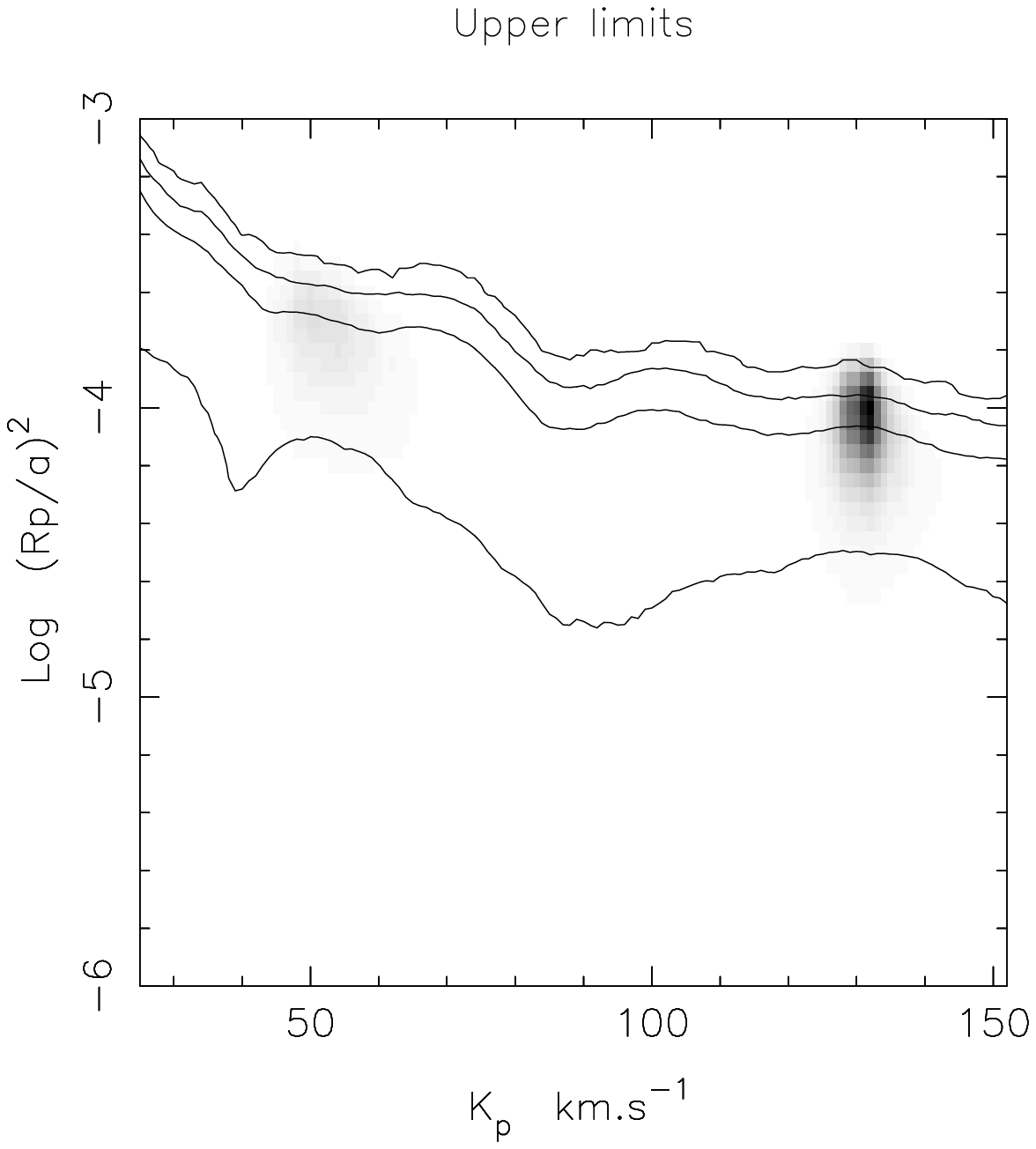,bbllx=106pt,bblly=73pt,bburx=481pt,bbury=404pt,angle=-90,width=8.6cm}
\caption{Relative probability map of model parameters $K_{p}$ and
$\log(\epsilon_{0}/p)=\log(R_{p}/a)^{2}$, derived from the WHT/UES
observations of $\upsilon$~And, assuming the albedo spectrum to be
that of a Class V roaster.  The greyscale and contours are defined as
in Fig.~\ref{fig:limits_obsgrey}.  }
\label{fig:limits_obs_V_mc}
\end{figure}

For comparison, we show in Fig~\ref{fig:limits_difgrey} a probability
map derived by subtracting Fig.~\ref{fig:ph_obsgrey} from
Fig.~\ref{fig:ph_sim80} to isolate the injected planet signal, then
performing the matched-filter analysis.  The injected planet is
clearly detected as a compact, dark feature with its correct amplitude
$\log(R_{p}/a)^{2}=2.57\times 10^{-4}$ and orbital velocity amplitude
velocity $K_{p}=137$ km~s$^{-1}$.  This most probable combination of
orbital velocity and planet radius yields an improvement
$\Delta\chi^{2}=98.6$ with respect to the value obtained assuming no
planet is present (i.e. $R_{p}=0$).  This is far greater than the
greatest $\Delta\chi^{2}=55.3$ produced at any value of $K_{p}$ in any
of the bootstrap trials.  The false-alarm probability for this
simulated signal is substantially less than one part in 3000, and the
``detection'' is secure.

We note that the calibration of the $(R_{p}/a)^{2}$ scale in
Fig.~\ref{fig:limits_obsgrey} depends on the value $p=0.42$ assumed
for the geometric albedo.  In the next sections, we explore the
relative ability of non-grey albedo models to fit the data.

\subsection{Class V roaster model}

The ``Class V roaster'' is the most highly reflective of the models
computed by \scite{sudarsky2000albedo}.  This model is characteristic
of planets with $T_{eff}\ge 1500$ K and/or surface gravities lower
than $\sim 10$ m s$^{-2}$, and has a silicate cloud deck located high
enough in the atmosphere that the overlying column density of gaseous
alkali metals is low, allowing a substantial fraction of incoming
photons at most optical wavelengths to be scattered back into space. 
There remains, however, a substantial absorption feature around the Na
I D lines, as shown in Fig.~\ref{fig:albsgeom}.  If $\upsilon$~And~b
has a mass and radius close to the values at the peak of the prior
probability distribution, its surface gravity is close to the critical
limit (Fig.~\ref{fig:grav_kp}).

We carried out the deconvolution using the same line list as for the
grey model, but with the line strengths attenuated using the Class V
albedo spectrum (see Appendix~\ref{sec:lsd}).  We back-projected the
resulting time series of deconvolved profiles
(Fig.~\ref{fig:ph_obs_V}) as described above.  We calibrated the
signal strength as described in Appendix~\ref{sec:calib}, by injecting
an artificial planet signature consisting of the spectrum of
$\upsilon$~And, attenuated by the Class V albedo spectrum and scaled
to the signal strength expected for a planet with $R_{p}=2 R_{jup}$. 
For the observations we used $\Delta v_{p}=8$ km~s$^{-1}$ which again
yielded the best fit, with scale factors and improvements in
$\chi^{2}$ as plotted in Fig.~\ref{fig:slice_V}.  The probability map
for the observed signals is shown in Fig.~\ref{fig:limits_obs_V_mc}.

The form of the Class V probability map is similar to that encountered
for the grey albedo spectrum.  The resulting upper limits on the
planet radius are listed in Table~\ref{tab:upperlim}.  The local
probability maxima near $K_{p}=52$ and 132 km~s$^{-1}$ are also
present with this albedo model.  As in the grey case, the feature at
$K_{p}=132$ is the most probable, this time by a slightly wider
margin.  Their signal strengths and false-alarm probabilities are
given in Table~\ref{tab:fap}.

\begin{figure}
\psfig{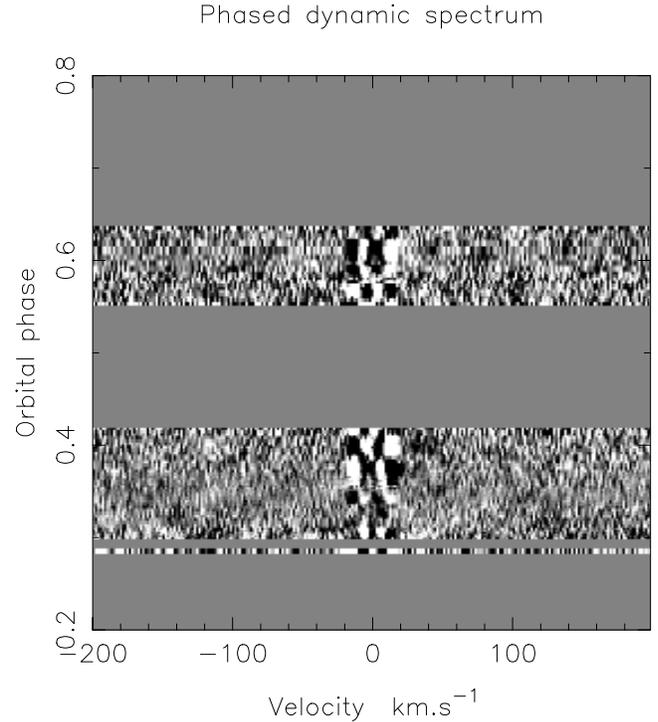} 
\caption[]{As for Fig.~\ref{fig:ph_obsgrey}, showing the time-series
of deconvolved profiles derived from the original observations 
assuming the albedo spectrum to be that of an ``isolated'' Class IV 
roaster.}
\label{fig:ph_obs_IVa}
\end{figure}

\begin{figure}
\psfig{figure=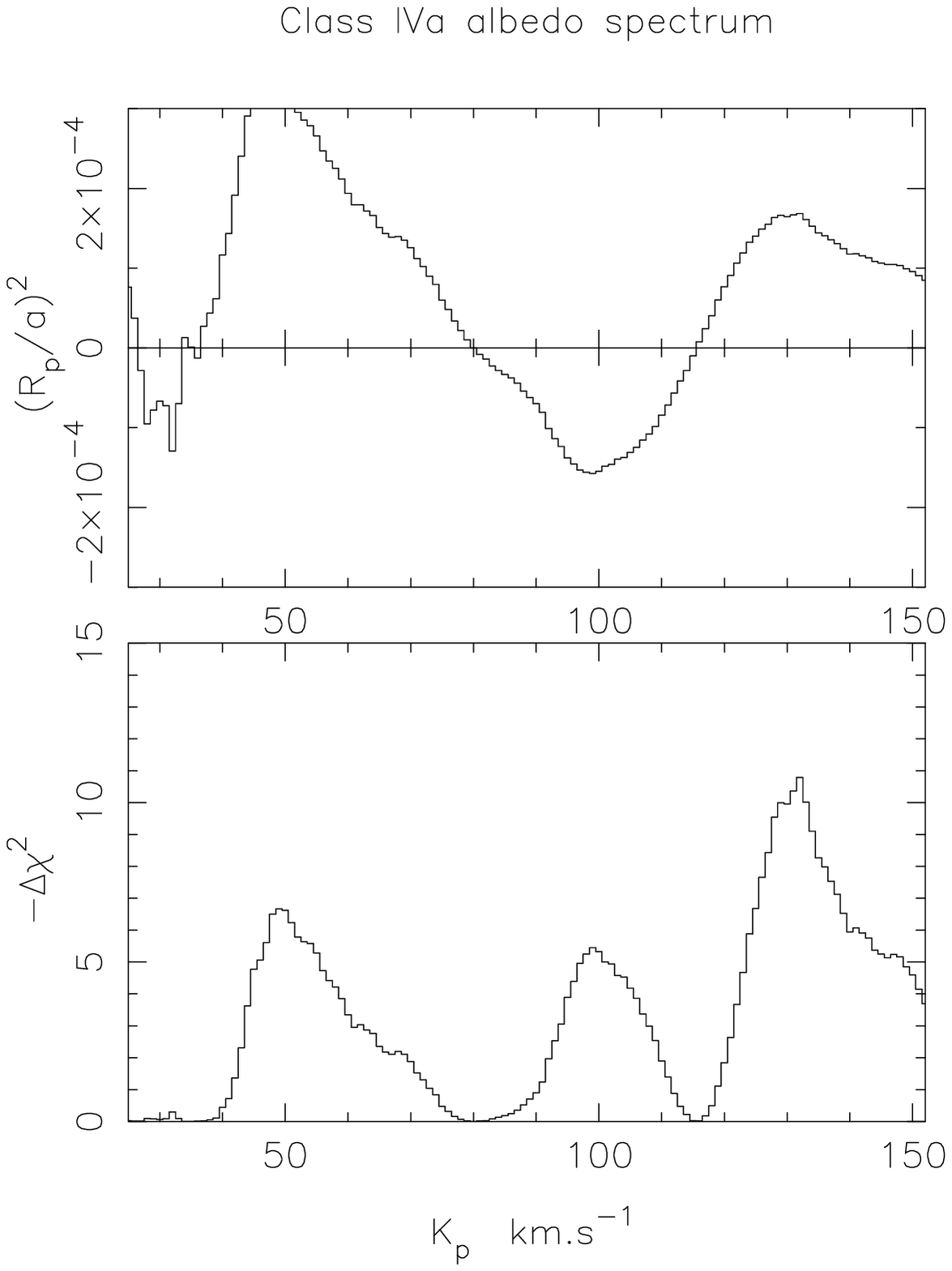,bbllx=7pt,bblly=0pt,bburx=350pt,bbury=453pt,width=8.5cm}
\caption{As for Fig.\ref{fig:slice_grey}, showing the 
optimal scaling factor 
$(R_{p}/a)^{2}$ and the associated
improvement in $\chi^{2}$ plotted against orbital 
velocity amplitude $K_{p}$, assuming an isolated Class IV albedo spectrum.
}
\label{fig:slice_IVa}
\end{figure}

\begin{figure}
\psfig{figure=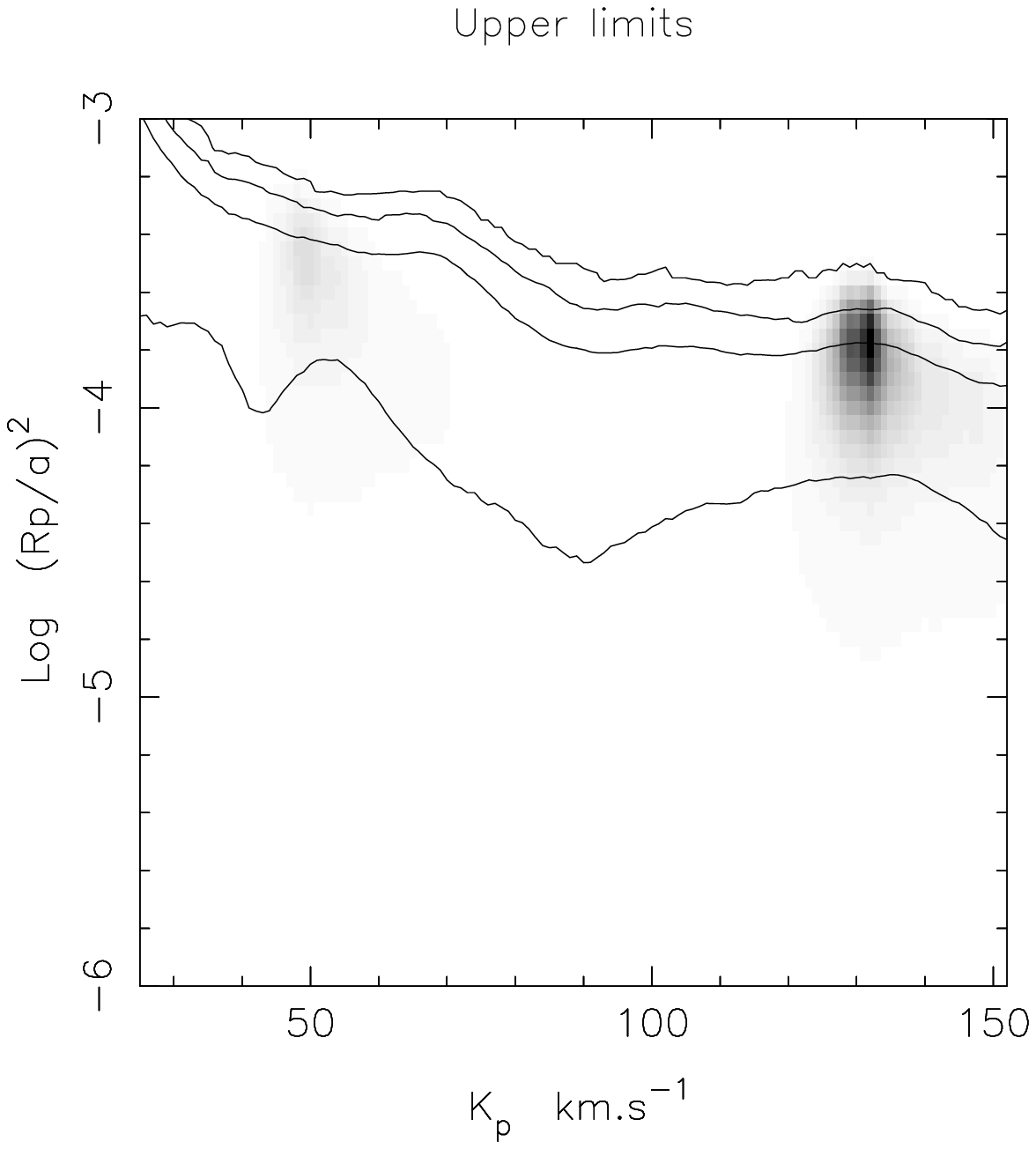,bbllx=106pt,bblly=73pt,bburx=481pt,bbury=404pt,angle=-90,width=8.6cm}
\caption{Relative probability map of model parameters $K_{p}$ and
$\log(\epsilon_{0}/p)=\log(R_{p}/a)^{2}$, derived from the WHT/UES
observations of $\upsilon$~And, assuming the albedo spectrum to be
that of an ``isolated'' Class IV roaster.  The greyscale and contours
are defined as in Fig.~\ref{fig:limits_obsgrey}.  }
\label{fig:limits_obs_IVa_mc}
\end{figure}

\subsection{Isolated Class IV model}

The Class IV roaster models of \scite{sudarsky2000albedo} have a more
deeply-buried cloud deck than the Class V models.  The resonance lines
of Na I and KI are strongly saturated, with broad wings due to
collisions with H$_{2}$ extending over much of the optical spectrum
(Fig.~\ref{fig:albsgeom}).

We used the procedures described above to deconvolve and back-project
the data assuming an ``isolated'' Class IV spectrum.  Although this
model does not take full account of the effects of irradiation of the
atmospheric temperature-pressure structure, it is a useful compromise
between the Class V models and the very low albedos found with
irradiated Class IV models.  The resulting time-series of deconvolved
spectra (Fig.~\ref{fig:ph_obs_IVa}) is noisier than the Class V and
grey-albedo versions, because lines redward of 500 nm contribute
little to the deconvolution. Consequently, the derived upper limits on 
the planet radius (Table~\ref{tab:upperlim}) are higher than for the 
Class V albedo model. 

The plots of $(R_{p}/a)^{2}$ and $\Delta\chi^{2}$ versus $K_{p}$ show
the same mix of positive and negative signal amplitudes encountered
with the grey and class V spectral models.  The back-projected
probability map (Fig.\ref{fig:limits_obs_IVa_mc}) nonetheless shows
the $K_{p}=132$ km~s$^{-1}$ probability maximum to be present and
again stronger than the 49 km/sec feature.  The fit to the data is,
however, poorer than in the grey and Class V cases, giving an
improvement of only $\Delta\chi^{2}=10.8$ over the no-planet
hypothesis and a planetary radius $R_{p}=1.68\pm 0.25 R_{Jup}$
(Table~\ref{tab:fap}).  This is 25\% larger than the radius estimated
with the Class V albedo model.  The bootstrap test gives a false-alarm
probability of 10\%.

This relatively large radius and poor fit appear to arise from a
mismatch between the wavelength dependence of the putative planet
signal and the Class IV model.  We tested this notion by deconvolving
and back-projecting a synthetic Class V model using the Class IV line
weights and flux calibration.  The radius of the planet was
over-estimated by 27\% in this experiment.  This occurs because the
least-squares deconvolution is attempting to fit a set of lines at all
wavelengths, with a deconvolution mask in which the lines at red
wavelengths are too weak to fit the data properly.  The best
compromise is achieved (in the least-squares sense) by over-estimating
the depth of the deconvolved profile in order to boost the strengths
of the weakened lines.

We conclude that the faint signals that contribute to the $K_{p}=132$
km~s$^{-1}$ feature are distributed in wavelength in a way that
resembles more closely a grey or a Class V spectrum than a
strongly-absorbed Class IV spectrum in which only blue light is
present.

\subsection{Plausibility of candidate signals}

Comparison of Figs.\ref{fig:limits_obsgrey}, \ref{fig:limits_obs_V_mc}
and \ref{fig:limits_obs_IVa_mc} with the joint prior probability
distribution in Fig.~\ref{fig:prior} shows that the feature near
$K_{p}=50$ km~s$^{-1}$ yields a value of $(R_{p}/a)^{2}$ that is
considerably greater than the radius predicted by
\scite{guillot97abos}.  The feature at $K_{p}=132\pm 3$ km~s$^{-1}$,
however, has $\epsilon_{0}=4.6\times 10^{-5}$.  A plausible geometric
albedo $p=0.42$ then gives $(R_{p}/a)^{2}=(1.09\pm 0.30)\times
10^{-4}$, which places it very close to the peak of the joint prior
probability distribution.

We probed the rotational broadening of the candidate features by
performing a series of backprojections using values of the gaussian
width parameter in the range $6.4 < \Delta v_{p} < 10$ km~s$^{-1}$,
corresponding to $ 0 < v_{e,refl} < 11.3$ km~s$^{-1}$.  The feature at
132 km~s$^{-1}$ yields the greatest improvement in $\chi^{2}$ when
$\Delta v_{p}=8\pm 1$ km~s$^{-1}$, or $v_{e,refl} =7.0\pm 2$ km
s$^{-1}$.  The 52 km~s$^{-1}$ feature, however, grows less significant
as $v_{e,refl}$ is decreased to the (retrograde) 5 km~s$^{-1}$ or so
expected at the corresponding orbital inclination.  The rotational
broadening of the 132 km~s$^{-1}$ feature is therefore consistent with
the value predicted in Fig.~\ref{fig:verefl_kp}.

We explored the region of parameter space around both features, by
carrying out back-projections on a grid of orbital periods and epochs
of zero phase around the values given in Table~\ref{tab:params}, taking
the epoch of transit to be near the middle of the data train to ensure
that the period and epoch were uncorrelated.  The local minimum in
$\chi^{2}$ was found to be centred at $K_{p}=133\pm 3$ km~s$^{-1}$,
$P=4.621\pm 0.005$ d, and $T_{0}=2451853.770\pm 0.025$.  The
radial-velocity ephemeris gives $P=4.61707 \pm 0.00003$ d and
$T_{0}=2451853.777\pm 0.012$, well within the 1$\sigma$ error bars
derived from the back-projection.  The 52 km~s$^{-1}$ feature, on the
other hand, is part of a more extended structure whose peak is located
at $K_{p}=67\pm 3$ km~s$^{-1}$, $P=4.651\pm 0.006$ d, and
$T_{0}=2451853.81\pm 0.02$.  This bears little relation to the
radial-velocity solution, suggesting a spurious origin.

We conclude that the candidate feature at $K_{p}=132$ km~s$^{-1}$
corresponds to a local minimum of $\chi^{2}$ with respect to the
parameters of interest, whose rotational broadening, phasing and
velocity amplitude are consistent with those expected of a
reflected-light signature from a planet whose orbital inclination is
near the most probable value. We cannot easily dismiss this feature 
as being merely an extension of a larger, spurious noise-induced 
structure.

\section{Conclusions}

The observations of $\upsilon$~And presented here rule out radii for
the innermost planet greater than $R_{p} > 0.98 R_{jup}/\sqrt{p}$ with
0.1\%\ false-alarm probability if a grey albedo spectrum is assumed.  We 
derive upper limits also with 0.1\%\ false-alarm probability
on the planet's radius of 1.53 $R_{jup}$
assuming the albedo spectrum of a Class V roaster (low gravity, high
cloud deck, \pcite{sudarsky2000albedo}), or 2.23 $R_{jup}$ for an
isolated Class IV model with saturated Na I D absorption from 550 nm
to the red limit of our spectra.  We cannot, however, rule out the
possibility that the planet has an albedo spectrum as bright as a
Class V roaster \cite{sudarsky2000albedo} if its radius is comparable
to that of HD 209458 b.  The evidence for a reflected-light signature
in the observations reported here, at a projected orbital velocity
amplitude $K_{p}=132\pm 3$ km~s$^{-1}$, is marginal but encouraging. 
If the orbital inclination is as close to edge-on as we infer from
previously-measured system parameters, the mass and radius of the
planet yield a surface gravity low enough for a Class V atmosphere to
be plausible.  If future observations confirm this signal, it gives a
radius $R_{p}=1.34\pm 0.17 R_{Jup}$ if a Class V spectrum is assumed. 
This is comparable to the radius of HD 209458 b inferred from {\em
HST} transit photometry \cite{charbonneau2000transit}.

We have explored the spectral properties of the candidate detection
via the relative probabilities of the fits to the data.  The Class V
roaster model gives a slightly better fit to the data than a grey
albedo model.  Both yield planet radii and orbital velocity amplitudes
close to the peak of the prior probablility density function.  The
isolated Class IV albedo model gives a significantly poorer fit to the
data.

While the possible detection discussed here gives believable physical
properties for the innermost planet, it remains too marginal for us to
claim a {\em bona fide} detection of reflected starlight.  Our
bootstrap analysis suggests a 7\%\ to 10\%\ likelihood that the
feature could be a spurious noise feature.  There is clearly a strong
case for a deeper reflected-light search to be made in the $\upsilon$
And system.

\section*{ACKNOWLEDGMENTS}

We thank David Sudarsky and Adam Burrows for providing us with
listings of their Class IV and Class V albedo models, and Geoff Marcy
for supplying us with his most recent orbital ephemeris and radial
velocity data for $\upsilon$~And~b.  This paper is based on
observations made with the 4.2-m William Herschel Telescope operated
on the island of La Palma by the Isaac Newton Group in the Spanish
Observatorio del Roque de los Muchachos of the Instituto de
Astrofisica de Canarias.  We are indebted to the support staff at the
ING for their assistance, and in particular Ian Skillen for doing the
tricky telescope scheduling, and John Telting for considerable effort
expended in implementing the telescope dithering procedure used in
these observations.  We acknowledge the support software and data
analysis facilities provided by the Starlink Project which is run by
CCLRC on behalf of PPARC. We thank the referee for raising several
useful points of clarification.


\appendix


\section{Removal of direct starlight} 
\label{sec:align}

\begin{figure*}
	\psfig{figure=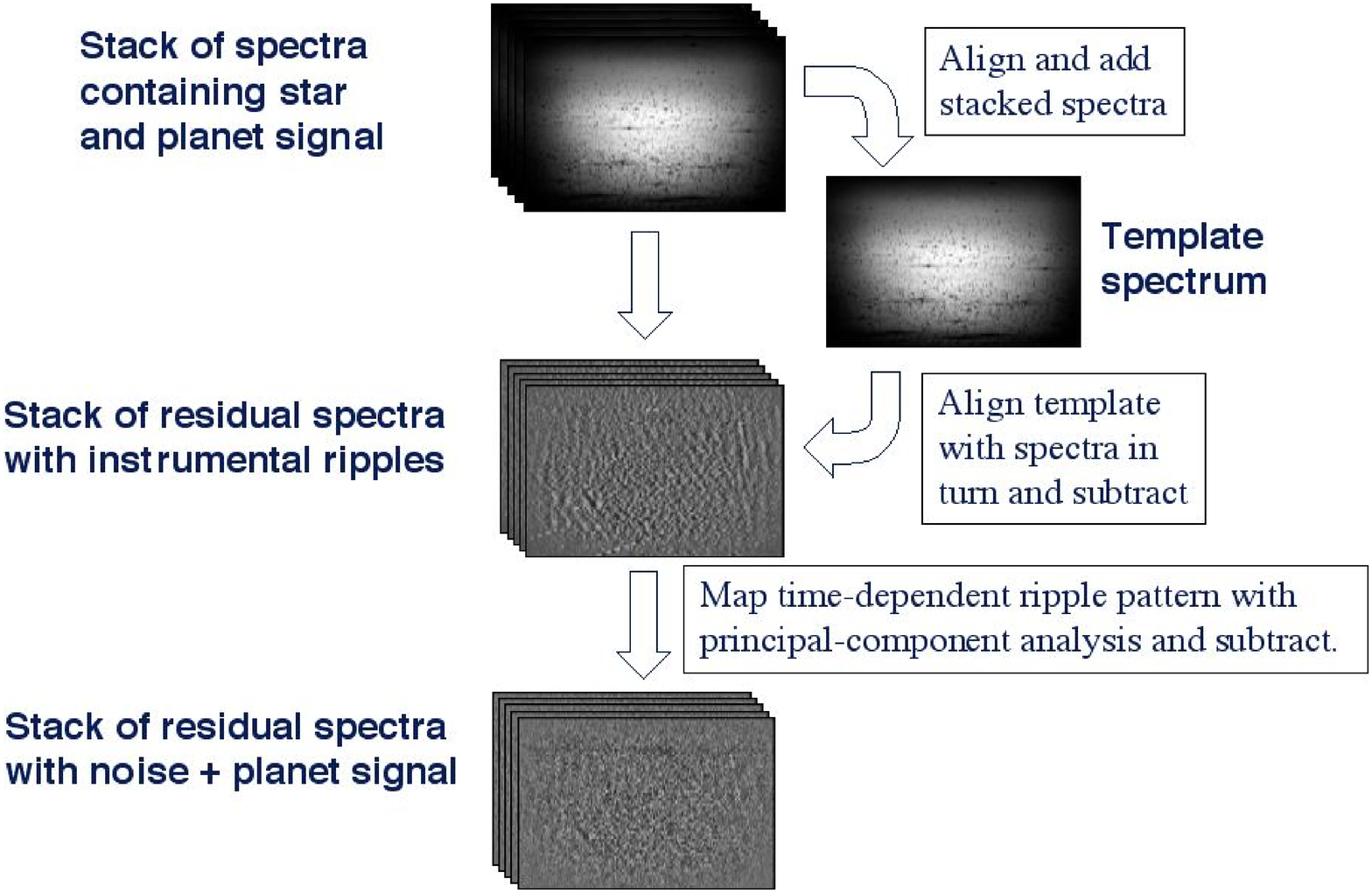,width=15cm}
	\caption[]{Schematic illustration of the main steps in 
	the construction and subtraction of the template spectrum, and the 
	subsequent fixed-pattern noise removal.}
	\label{fig:flow_align}
\end{figure*}

In this appendix we describe the procedure we use to subtract the
stellar spectrum without introducing substantial noise and also
without subtracting the planet signal.  The intrinsic spectrum of the
star was modelled as the co-aligned sum of all the spectra of
$\upsilon$~And secured during the run.

We began by removing cosmic-ray hits from the individual frames. 
Within each night's data, each frame was divided by its predecessor
and the result was median smoothed with a box covering 21 pixels in
the dispersion direction and 5 adjacent orders.  This smoothed frame
was used to scale the predecessor frame, which was then subtracted
from the frame under consideration.  The difference frame was divided
by the square root of the frame under consideration, to yield deviates
in units of the local Poissonian noise amplitude.  All pixels whose
deviates were more than 7 sigma in excess of their counterparts in the
scaled predecessor frame were assumed to be cosmic ray hits.  Their
values were replaced with values taken from the corresponding pixels
in the scaled predecessor frame.  The result was re-scaled and added
to the scaled predecessor frame to restore the original frame with all
major cosmic-ray hits removed.  The cleaned frames were then shifted
to co-align the stellar absorption lines to the nearest integer pixel. 
Finally the cleaned frames were summed to make the template frame
(step 1 in Fig.~\ref{fig:flow_align}).

The individual cleaned frames were co-added in groups of four for the
next stage in the procedure, which involved scaling, shifting and
blurring (or sharpening) the template frame prior to subtraction from
each group.  The first and second derivatives of the template frame
were computed from the template values $t_{j}$ in each order as a
function of pixel number $x_{j}$ in the dispersion direction:
\begin{equation}
t'_{j} = \frac{t_{j+1}-t_{j-1}}{x_{j+1}-x_{j-1}}
\end{equation}
and
\begin{equation}
t''_{j}=2\frac{(t_{j+1}-t_{j})/(x_{j+1}-x_{j})-(t_{j}-t_{j-1})/(x_{j}-x_{j-1})}
{x_{j+1}-x_{j-1}}.
\end{equation}
The third and fourth derivatives were computed by applying the same 
operations to $\bvec{t}''$, for use in correcting frame-to-frame 
changes in the higher moments of the PSF and seeing profile.

The vector $\bvec{f}$ of all observed spectrum values within each 
extracted echelle order was then modelled by scaling the template 
$\bvec{t}$ along each order, the scale factors varying as a function 
of position $\bvec{x}$ approximated by a 34-knot least-squares spline.  
The use of such a high-order spline was demanded by vignetting near 
the edges of each order, which produced an abrupt and time-variable 
change in the slope of the stellar continuum towards the end of each 
order.  The derivative frames were scaled in a similar fashion, using 
6-knot splines to define the scale factors.  An additional 6-knot 
spline was added to the fit to provide a smooth background correction 
for any inconsistencies in the scattered-light subtraction during the 
extraction process.  The scaled, aligned, distorted template spectrum 
thus had the form
\begin{equation}
g_{j}=\alpha_{j}t_{j} + \beta_{j}t'_{j} + \gamma_{j}t''_{j} + 
\delta_{j}t'''_{j} + \epsilon_{j}t''''_{j} + \eta_{j}
\end{equation}
where $\alpha_{j}$ is the value at $x_{j}$ of the 34-knot spline and 
$\beta_{j}$, $\gamma_{j}$, $\delta_{j}$, $\epsilon_{j}$ and $\eta_{j}$ 
are the values of the corresponding 6-knot splines used to scale 
the derivatives and correct the background.  The knot values for the 
various splines were determined by the method of least squares using 
singular-value decomposition to ensure that spurious fluctuations were 
suppressed in those parts of the spectrum where few lines were 
present.

When the planet velocity is sufficient to shift the absorption lines
in the reflected starlight well away from those in the direct
starlight, then the residual spectrum $\bvec{r}=\bvec{f}-\bvec{g}$ is
expected to retain the reflected starlight signal, Doppler shifted and
deeply buried in Poissonian noise, once step 2 in 
Fig.~\ref{fig:flow_align} is complete.

\section{Residual fixed-pattern noise}
\label{sec:fpnoise}

In this appendix we describe the principal-component analysis method
used to correct the spatially fixed but time-dependent ripples found
in the residual spectra following subtraction of the template. 

Typically the grouped spectra contained $4\times 10^{6}$ or more
photons per pixel over most of the recorded spectrum.  The scatter of
the pixel values in the residual spectrum was compared with
expectations from photon statistics, by dividing each grouped spectrum
through by the square root of its computed variance.  The distribution
of the pixel deviates calculated over the 62 orders used for the
deconvolution (see Appendix~\ref{sec:lsd} below) was found typically
to contain a few hundred extreme outliers, mostly caused by systematic
errors in the polynomial fits in half a dozen regions affected by
defects on the CCD. Despite clipping at $\pm 4$ times the expected
local root-mean-square (rms) photon noise amplitude, we found that the
distribution of the remaining values was Gaussian with an rms error
between 1.4 and 1.6 times the expected value.  We found a significant
correlation between pixel values in successive difference frames,
indicating that some fixed-pattern noise sources had not been
eliminated fully by the template alignment and subtraction.  The
residuals were correlated over a few hours in time.  The residuals in
pairs of frames taken several hours apart or on different nights were
also found to be correlated, though in some instances the slope of the
correlation was reduced or even reversed.  The fixed-pattern noise is
visible as ripples in the example frame shown following step 2 in
Fig.~\ref{fig:flow_align}.

We mapped these spatially fixed but time-varying patterns in the
difference spectra using principal-component analysis (PCA).  We start
with a set of $M$ spectra, each of length $N$ pixels.  The $i$th
spectrum thus has elements
$\bvec{s}_{i}=\{s_{i1},s_{i2},\ldots,s_{iN}\}$ and associated
variances
$\bvec{\sigma^{2}}_{i}=\{\sigma^{2}_{i1},\sigma^{2}_{i2},\ldots,\sigma^{2}_{iN}\}$. 
The spatial correlation matrix of the time variations in the
individual spectral bins of this set of spectra is a real symmetric
matrix of dimensions $N\times N$, whose elements are given by:
$$
H_{jk} = 
\sum_{i=1}^{M}\frac{(s_{ij}-\hat{s}_{j})(s_{ik}-\hat{s}_{k})}
            {\sigma_{ij}\sigma_{ik}}.
$$
Note that we subtract the inverse variance weighted mean spectrum
$\hat{s}$ before computing the correlation matrix. The $j$th
spectral bin of $\hat{s}$ is given by:
$$
\hat{s}_{j} = \frac{\sum_{i=1}^{M} s_{ij}/\sigma^{2}_{ij}}
                   {\sum_{i=1}^{M} 1/\sigma^{2}_{ij}}.
$$
The principal components are those eigenvectors of the correlation
matrix accounting for the largest fraction of the variance, and in our
implementation are computed using Jacobi's method
\cite{press92numrec}.  To keep computing time down, we processed the
spectra in segments of length 100 pixels.

We found that most of the unwanted additional variance was contained
in the first two principal components, indicating that two independent
sources of fixed-pattern noise were present.  The first principal
component contained a spatially fixed but temporally-varying pattern
of ripples affecting all orders, which is probably attributable to a
slow drift in the sensitivity pattern of the flat field and/or
thermal flexure in the spectograph.  The second
principal component consisted mainly of imperfectly-subtracted
telluric absorption features in some orders.  We computed the
contributions of these two fixed-pattern noise sources to each
exposure and subtracted them from the difference frames.  This
effectively removed the correlation between successive frames, and
reduced the RMS scatter in the pixel values to within $\pm 10$\%\ of
the value expected from photon statistics alone (step 3 in
Fig.~\ref{fig:flow_align}).  The planet signature should not be
affected by this procedure any more than it is affected by the
template subtraction because, unlike the fixed-pattern noise, it is
smeared out by orbital motion.

\section{Least-squares deconvolution}
\label{sec:lsd}

\begin{figure*}
	\psfig{figure=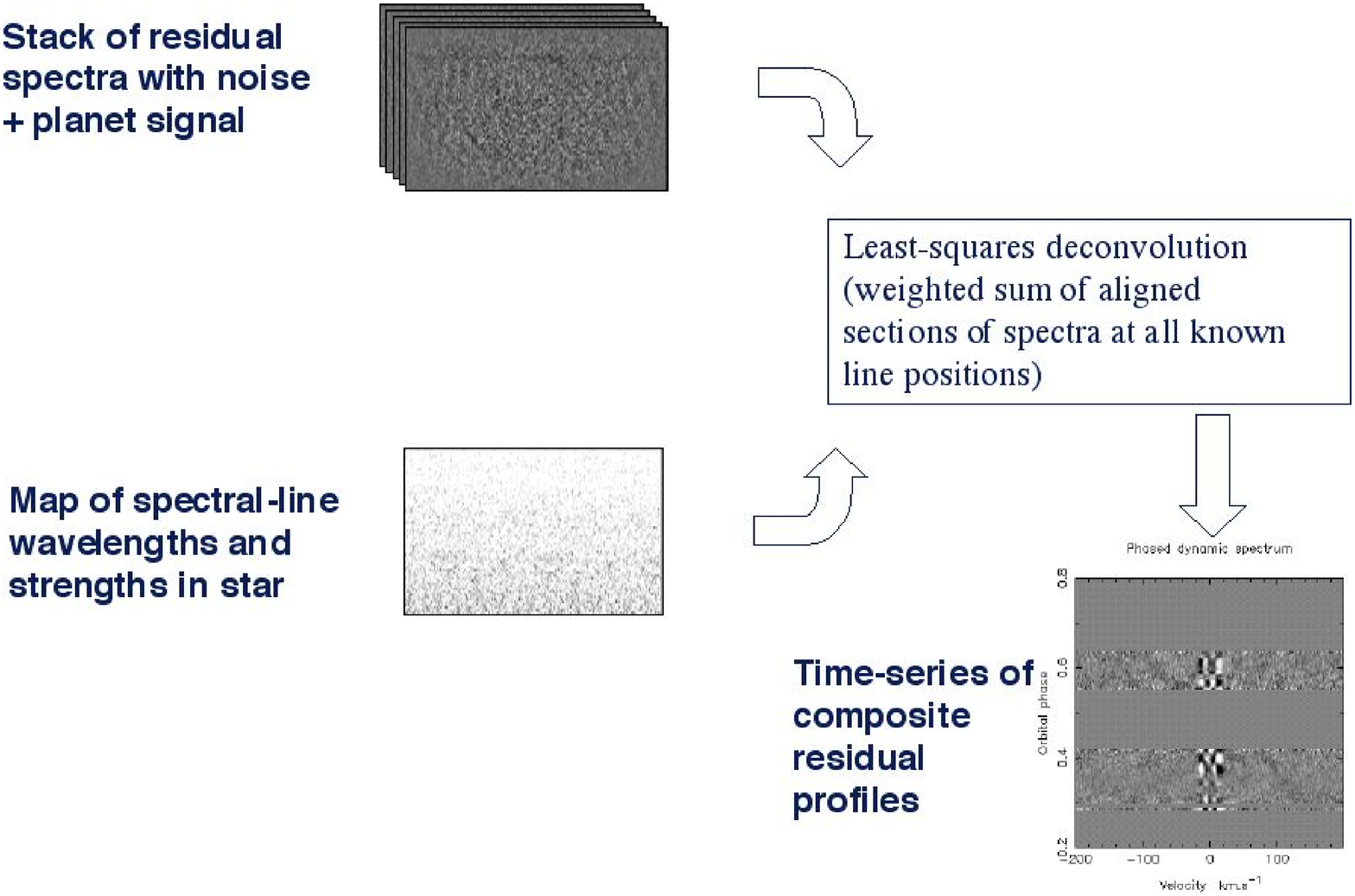,width=15cm}
	\caption[]{Schematic illustration of the main steps in 
	the least-squares deconvolution of a time-series of spectra.}
	\label{fig:flow_lsd}
\end{figure*}

We extracted the reflected-light signature from the difference frames
using the method of weighted least-squares deconvolution (LSD)
described by \scite{donati97zdi}.  This method, shown schematically in
Fig.~\ref{fig:flow_lsd}) entails taking a list of lines with relative
strengths appropriate to the type of star concerned, and computing via
the method of least squares a ``mean'' line profile which, when
convolved with the line pattern, gives an optimal match to the
observed spectrum.  The deconvolved profile thus incorporates an
average broadening function that is representative of all the lines
recorded in the spectrum.  Applied to the residual spectrum from which
the direct stellar signal has been removed, LSD is an effective way of
measuring the average line profile of the faint reflected-light
signature of the planet, because the reflected light should should
have the same pattern of absorption-line positions and strengths as
the stellar spectrum.

In terms of signal improvement, LSD is analogous to aligning and
averaging (with appropriate weighting factors for line strength and
local continuum signal strength on the recorded frame) the profiles of
all the individual photospheric absorption lines recorded on each
echellogram and included in the line list.  As each individual
spectral line appears in at least two adjacent echelle orders, we have
7700 images of the 3450 spectral lines listed in the observed
wavelength range.  If all lines were of equal strength and the
continuum signal were constant over the whole frame, the
signal-to-noise deconvolved profile would be proportional to the
square root of the number of line images used, i.e. nearly 70 times
greater than the signal-to-noise ratio of a single line in the
original spectrum.  In practice, the recorded continuum is not uniform
and the lines have a wide variety of depths.  Even so, the
signal-to-noise ratio of the deconvolved profile is found to be some
30 times greater than that of a single line in the best-exposed parts
of the original spectrum.  The least-squares fitting procedure has the
additional advantage that neighbouring, blended lines are treated
simultaneously, thereby eliminating the sidelobes that would be
produced by a simpler shift-and-add procedure.

We divide the observed residual spectrum by the continuum fit to the
original spectrum, to obtain a normalised residual spectrum $\bvec{r}$
expressed in units of the continuum level of the original spectrum. 
We modify the variances associated with the elements of $\bvec{r}$
accordingly.  We treat $\bvec{r}$ as the convolution of a ``mean''
line profile $\bvec{z}(v)$ with a set of weighted delta functions at
the wavelengths of a comprehensive list of spectral lines.  The
profile $\bvec{z}$ is defined on a linear velocity scale, with a
velocity increment $\Delta v=3$ km~s$^{-1}$ per bin, which is close to
the average velocity increment per pixel in the extracted spectra. 
The elements $A_{jk}$ of the convolution matrix $A$ are computed by
summing, over all spectral lines $\ell$, the fractional contribution of
the element of the deconvolved profile at velocity $v_{k}$ to the data
pixel at wavelength $\lambda_{j}$ when the centre of the deconvolved
profile is shifted to the wavelength $\lambda_{\ell}$ of each line in
turn.  Hence
\begin{equation}
A_{jk}=\sum_{\ell}w_{\ell}\Lambda[(v_{k}-c(\lambda_{j}/\lambda_{\ell}-1))/\Delta 
v].
\end{equation}
The triangular function $\Lambda$ has the form $\Lambda(x)=1+x$ for
$-1<x\le 0$, $\Lambda(x)=1-x$ for $0<x< 1$, and is zero everywhere
else.  The line weights $w_{\ell}$, incorporated in the convolution
matrix $\bvec{A}$, are proportional to the central depths of the lines
as computed from a Kurucz model atmosphere for the appropriate
spectral type.  After some experimentation we found that a line list
computed for a G2 spectral type with solar abundances gave the best
results.  The use of a cooler template gives a better fit to the line
depths than an earlier spectral type, presumably because of the star's
above-solar metallicity.  

Because the form of the geometric albedo spectrum of $\upsilon$~And~b
is unknown, we investigated the effects of both grey (i.e.
wavelength-independent) and non-grey geometric albedo spectra.  The
non-grey models we employed were the Class V and ``isolated'' Class IV
models of \scite{sudarsky2000albedo} (Fig.~\ref{fig:albsgeom}).  The
theoretical albedo spectra provided an additional weighting factor in
the deconvolution linelist, allowing us to place limits on the
planet's radius for any assumed albedo model, and to assess the
relative goodness-of-fit to the data for different atmospheric models.
 
The deconvolved profile $\bvec{z}(v)$ has the form of a line profile 
normalized to unit continuum intensity, but from which the continuum 
has been subtracted.  Determination of $\bvec{z}(v)$ via least squares 
entails minimizing the magnitude of the misfit vector 
$|\bvec{r}-\bvec{A}\cdot\bvec{z}|$,
\begin{equation}
\chi^2=(\bvec{r}-\bvec{A}\cdot\bvec{z})^T\cdot\bvec{V}\cdot(\bvec{r}-\bvec{A}\cdot\bvec{z}),
\end{equation}
weighted so as to make due allowance for the observational errors 
$\sigma_{j}$ associated with the individual spectral bins.  Here the 
inverse variances $\sigma^{2}_{j}$ associated with the $N$ elements 
$r_{j}$ of the residual spectrum $\bvec{r}$ are incorporated via the 
diagonal matrix
\begin{equation}
\bvec{V}=\mbox{Diag}[1/\sigma^{2}_{1},\ldots,1/\sigma^{2}_{N}].
\label{eq:invar}
\end{equation}
The least-squares solution for $\bvec{z}$ is found by solving the 
matrix equation
\begin{equation}
\bvec{A}^{T}\cdot\bvec{V}\cdot\bvec{A}\cdot\bvec{z} = \bvec{A}^{T}\cdot\bvec{V}\cdot\bvec{r}.
\label{eq:deconvol}
\end{equation}
Since the square matrix $\bvec{A}^{T}\cdot\bvec{V}.\bvec{A}$ is symmetric 
and positive-definite, the least-squares problem can be solved using 
efficient methods such as Cholesky decomposition \cite{press92numrec}.

The deconvolved profile $\bvec{z}$ is expressed in units of the 
weighted mean continuum level.  The deconvolution procedure 
compensates for local line blends, and so has the advantage over 
cross-correlation methods that outside the region occupied by the 
residual stellar profile, the deconvolved spectrum is flat.  The 
formal errors on the $M$ points of the deconvolved profile $\bvec{z}$ 
are obtained in the usual way from the diagonal elements of the 
$M\times M$ covariance matrix
\begin{equation}
\bvec{C}=[\bvec{A}^{T}\cdot\bvec{V}\cdot\bvec{A}]^{-1}.
\end{equation}


\section{Matched-filter analysis}
\label{sec:matchfilt}

\begin{figure*}
	\psfig{figure=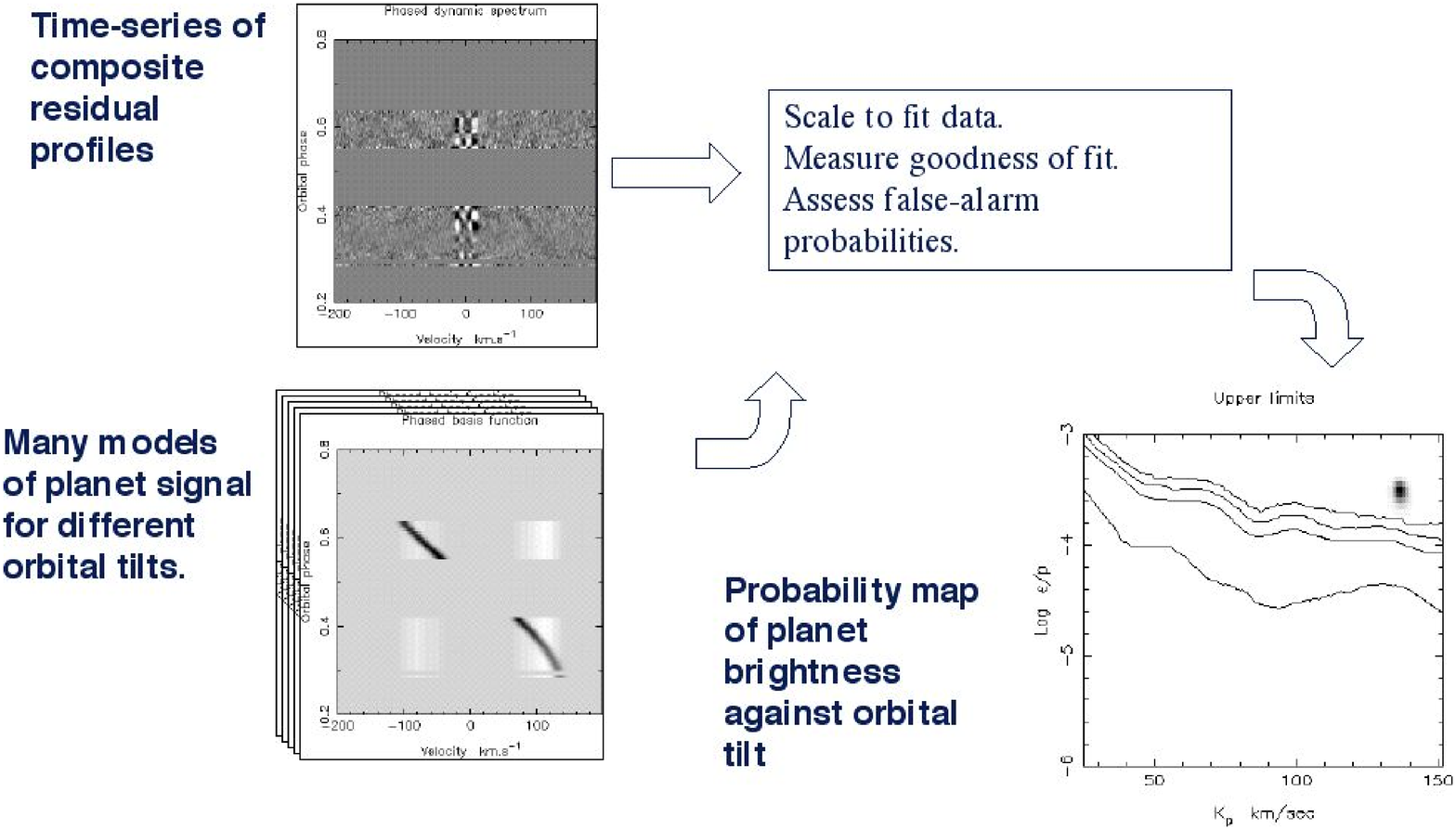,width=15cm}
	\caption[]{Schematic illustration of the main steps in 
	the matched-filter analysis.}
	\label{fig:flow_matchfilt}
\end{figure*}

To detect the planet we must co-align and add its signal from profiles 
at many different orbit phases.  Given an assumed orbital inclination, 
we can compute the sinusoidal path that the planet should follow 
through velocity space, together with the attendant changes in signal 
strength. To search for a pattern of faint features displaying the 
expected behaviour, we construct a set of matched filters that can be scaled 
to fit the data for each member of a set of appropriate orbital 
inclinations, as shown schematically in Fig.~\ref{fig:flow_matchfilt}. 

We model the velocity profile of the reflected-light signal as
a sequence of gaussians with appropriate velocities and 
relative amplitudes:
\begin{eqnarray}
G(v,\phi | K_{p}) &=& \frac{W_{\star}}{\Delta v_{p}\sqrt{2\pi}}
g(\phi,i)\times\nonumber\\
&&\exp\left[-\frac{1}{2}
\left(\frac{v-K_{p}\sin\phi}{\Delta v_{p}}\right)^{2}\right].
\label{eq:basfunc}
\end{eqnarray}
The amplitude $K_{p}$ of the sinusoidal velocity variation is
determined by the system inclination and stellar mass according to
\begin{equation}
	K_{p} \simeq 139
       \left( \frac{ M_\star }{ 1.3 M_\odot } \right)^{1/3}
	    \sin{i} 
       {\rm km~s}^{-1}.
\end{equation}

The phase angle $\alpha$ can be determined at any orbital phase from
\begin{equation}
\cos\alpha = - \sin i .\cos 2\pi\phi.
\end{equation}

$W_{\star}$ is the integrated area of the deconvolved
stellar line profile, and $\Delta v_{p}$ is the characteristic width
of the planet's reflected-light profile.  The width $\Delta v_{p}=9.1
$ km~s$^{-1}$ of the gaussian representing the planetary profile was
determined from a fit to the deconvolved profile of $\upsilon$~And. 

\subsection{Phase function}
\label{sec:phasefuncs}

For the phase function there are two natural choices. First, the 
phase function of a Lambert sphere is
\begin{equation}
g(\alpha)=(\sin\alpha+(\pi-\alpha)\cos\alpha)/\pi.
\label{eq:g_lambert}
\end{equation}
This assumes that the planetary atmosphere scatters isotropically into 
$2\pi$ steradians. 

Second, it may be more realistic to adopt a phase function that
resembles those for the cloud-covered surfaces of planets in our own
solar system.  Jupiter and Venus appear to have phase functions that
are more strongly back-scattering than a Lambert sphere.  Photometric
studies of Jupiter at large phase angles from the {\em Pioneer} and
subsequent missions have shown \cite{hovenier89} that the phase
function for Jupiter is very similar to that of Venus.  As a plausible
alternative to the Lambert-sphere formulation, we use a polynomial
approximation to the empirically determined phase function for Venus
\cite{hilton92}.  The phase-dependent correction to the planet's
visual magnitude is approximated by:
\begin{equation}
\Delta m(\alpha)=0.09(\alpha/100^\circ) + 2.39(\alpha/100^\circ)^2 
- 0.65(\alpha/100^\circ)^3.
\label{eq:dm_venus}
\end{equation}
so that
\begin{equation}
g(\alpha)=10^{-0.4 \Delta m(\alpha)}.
\label{eq:g_venus}
\end{equation}

\subsection{Attenuation during starlight subtraction}

The template spectrum that is subtracted in turn from each spectrum
contains a blurred planet signal, as described in
Appendix~\ref{sec:align}.  Our matched filter must therefore mimic
accurately the effects of subtracting a template constructed from the
spectra themselves, so we subtract the inverse variance-weighted average
of the travelling gaussian.  If $\sigma^{2}_{ij}$ is the variance of
the $i$th velocity bin in the $j$th deconvolved profile, then the
attenuated basis function becomes:
\begin{equation}
	h_{ij}(K_{p})=G(v_{i},\phi_{j} | K_{p}) -
	\frac{\sum_{j}G(v_{i},\phi_{j} | K_{p})/\sigma^{2}_{ij}}
	{\sum_{j}1/\sigma^{2}_{ij}}.
	\label{eq:attenuate}
\end{equation}

To allow for any systematic errors in the continuum level following
template subtraction and deconvolution, we subtract the inverse
variance-weighted mean value of each of the deconvolved profiles at
each orbital phase.  If $d_{ij}$ is the original data value at 
velocity $i$ and phase $j$, then
\begin{equation}
	D_{ij}=d_{ij}-\frac{\sum_{i}d_{ij}/\sigma^{2}_{ij}}{1/\sigma^{2}_{ij}}
\end{equation}
gives the resulting orthogonalised data pixel value. The matched 
filter is orthogonalised in the same way:
\begin{equation}
	H_{ij}=h_{ij}-\frac{\sum_{i}h_{ij}/\sigma^{2}_{ij}}{1/\sigma^{2}_{ij}}.
\end{equation}

\subsection{Scaling the matched filter}

The attenuated basis function $H(v,\phi,K_{p})$ is normalised such 
that when it is scaled to give an optimal fit to the entire time-series 
of $D_{ij}$ values, the scaling factor is $\epsilon_{0}$, the
planet/star flux ratio at $\alpha=0$ as defined in eq.~\ref{eq:eps0}. 

The optimal scale factor
$\epsilon_{0}$ and its formal error are determined from
\begin{equation}
\epsilon_{0}(K_{p})=\sum_{i,j}
\frac{D_{ij}H_{ij}(K_{p})/\sigma^{2}_{ij}}
{H_{ij}^{2}(v_{i},\phi_{j},K_{p})/\sigma^{2}_{ij}}
\label{eq:fiteps}
\end{equation}
and
\begin{equation}
\mbox{Var}(\epsilon_{0})=\sum_{i,j}
\frac{1}
{H_{ij}^{2}(K_{p})/\sigma^{2}_{ij}}.
\label{eq:vareps}
\end{equation}

We exclude all pixels within 25 km~s$^{-1}$ of the stellar line core 
from the summations, to eliminate spurious effects arising from the ripple 
pattern in the core of the residual deconvolved stellar profile.
 
\subsection{Calibrating non-grey albedo models.}
\label{sec:calib}

The meaning of $\epsilon_{0}$ in the expressions above is obvious if
the albedo spectrum is grey, i.e. if $\epsilon_{0}$ is independent of
wavelength.  Problems arise, however, when we wish to test how well a
given non-grey albedo spectrum fits the data.  In this case, the
weighting factors $w_{i}$ used for the lines in the deconvolution mask
mask are multiplied by the albedo as described in
Appendix~\ref{sec:lsd} above, and $\epsilon_{0}$ follows the
wavelength dependence of the geometric albedo.

We circumvented this difficulty by replacing $\epsilon_{0}$ with
$(R_{p}/a)^{2}$ as the scaling factor in eq.~\ref{eq:fiteps}. 

The basis function $G$ is rescaled to become $G'=\left< p \right>G$,
where $\left< p \right>$ is an appropriately weighted
wavelength-averaged geometric albedo, then attenuated as before using
eq.~\ref{eq:attenuate}.  For each candidate albedo spectrum we
determine the appropriate value of $\left< p \right>$ empirically.  We
inject an artificial planet signal with the required albedo spectrum
and known $K_{p}$ and $R_{p}/a$ into the data, and deconvolve the
synthetic data with the albedo-weighted line list.  We subtract the
profiles of the observed spectra, following deconvolution with the
same line list, to isolate the deconvolved planet signal from the
noise.  We then back-project the isolated planet signal, and measure
the value of $\left< p \right>$ that recovers the correct value of
$R_{p}/a$ at the appropriate $K_{p}$.


\section{False-alarm probabilities for candidate signals}
\label{sec:fap}

Candidate reflected-light signals are characterised by positive values
of $(R_{p}/a)^{2}$ and an improvement in $\chi^{2}$ relative to the
fit obtained when $(R_{p}/a)^{2}=0$. We determine the frequency with 
which  $(R_{p}/a)^{2}$ exceeds a given value due to noise in the 
absence of a planet signal, using a bootstrap procedure. 

In each of 3000 trials, we randomize the order in which the three
nights of observations were secured, then we randomise the order in
which the observations were secured within each night.  The re-ordered
observations are then associated with the original sequence of dates
and times.  This ensures that any contiguous blocks of spectra
containing similar systematic errors remain together, but appear at a
new phase.  Any genuine planet signal present in the data is, however,
completely scrambled in phase.  The re-ordered data are therefore as
capable as the original data of producing spurious detections through
chance alignments of blocks of systematic errors along a single
sinusoidal path through the data.  We record the least-squares
estimates of $(R_{p}/a)^{2}$ and the associated values of $\chi^{2}$
as functions of $K_{p}$ in each trial.  The 3000 trials implicitly
define empirical probability distributions of $(R_{p}/a)^{2}$ and
$\chi^2$ that include both the photon statistics and the effects of
correlated systematic errors at each trial value of $K_{p}$.

The percentile points of the distribution of $(R_{p}/a)^{2}$ values at
each $K_{p}$ are used to define the 1$\sigma$, 2$\sigma$, 3$\sigma$
and 4$\sigma$ upper-limit contours shown in
Figs.~\ref{fig:limits_obsgrey}, \ref{fig:limits_difgrey},
\ref{fig:limits_obsgrey}, \ref{fig:limits_obs_V_mc} and
\ref{fig:limits_obs_IVa_mc}.

To assess the false-alarm probability for a candidate detection,
however, we need to examine the likelihood of the fit to the data.  In
each bootstrap trial, the most likely candidate is the one among those
with $(R_{p}/a)^{2}>0$ that yields the greatest improvement
$\Delta\chi^{2}$ relative to the no-planet hypothesis.  Such a peak
can occur at any value of $K_{p}$.  We use the distribution of the
peak $\Delta\chi^{2}$ value in each trial to determine how often we
would expect the $\Delta\chi^{2}$ of the strongest spurious noise
feature giving $(R_{p}/a)^{2}>0$ to exceed the actual $\Delta\chi^{2}$
of a candidate detection in the observed data.  We define this
probability -- summed over all possible values of $K_{p}$ with some
weighting according to prior probability estimates -- to be the
``false-alarm probability'' for a candidate signal.

The unweighted false-alarm probabilities are based on the likelihood 
of obtaining the data $\bvec{D}$ at a given $K_{p}$ after optimizing 
$(R_{p}/a)^{2}$, and take no account of whether or not that value of 
$K_{p}$ is plausible. This conditional likelihood is measured 
relative to the likelihood of getting the same data if no planet 
signal is present:
\begin{equation}
	\frac{P(\bvec{D}|K_{p})}{P(\bvec{D}|\mbox{No planet})} = 
	e^{-(\chi^{2}(K_{p})-\chi^{2}(\mbox{No planet}))/2},
\end{equation}
where we have implicitly optimised the value of $(R_{p}/a)^{2}$
using the matched-filter fit at each $K_{p}$. 

We take our prior knowledge of $K_{p}$ into account by modifying the
likelihood to give the joint probability
\begin{equation}
	P(\bvec{D},K_{p}) = P(\bvec{D}|K_{p}) P(K_{p}).
\end{equation}
We use the prior probability distribution for $K_{p}$ as shown
projected on to the $K_{p}$ axis in Fig.\ref{fig:prior}.

We compute $P(\bvec{D},K_{p})$
for each of the 3000 bootstrap trials.  To do this, we pick out the
greatest value in each trial of the quantity
\begin{equation}
	\ln P(\bvec{D},K_{p}) = \ln P(K_{p}) +\frac{\Delta\chi^{2}}{2}
\end{equation}
where $P(K_{p})$ is derived from the simulations described in
Section~\ref{sec:prior}.  The cumulative distribution of maximal $\log
P(\bvec{D},K_{p})$ values derived from the bootstrap trials gives the
probability that the observed peak value of $\ln P(\bvec{D},K_{p})$
could be spurious.  

\end{document}